\documentclass[apj,numberedappendix,twocolappendix]{emulateapj}
\usepackage{graphicx}
\usepackage{physics}
\usepackage{xcolor}
\usepackage{hyperref}

\begin{document}

\shorttitle{The 21 cm signal and the light cone effect}
\shortauthors{La Plante et al.}

\title{Reionization on Large Scales IV: \\ Predictions for the 21 cm signal incorporating the light cone effect}
\author{P. La Plante\altaffilmark{1}\altaffilmark{*}, N. Battaglia\altaffilmark{1}, A. Natarajan\altaffilmark{1,2}, J. B. Peterson\altaffilmark{1}, H. Trac\altaffilmark{1}, R. Cen\altaffilmark{3}, A. Loeb\altaffilmark{4}}

\altaffiltext{1}{McWilliams Center for Cosmology, Department of Physics, Carnegie Mellon University, Pittsburgh, PA 15213, USA}
\altaffiltext{2}{Department of Physics and Astronomy \& Pittsburgh Particle physics, Astrophysics and Cosmology Center, University of Pittsburgh, Pittsburgh, PA 15260, USA}
\altaffiltext{3}{Department of Astrophysical Science, Princeton University, Princeton, NJ 08544}
\altaffiltext{4}{Harvard-Smithsonian Center for Astrophysics, Cambridge, MA 02138}
\altaffiltext{*}{plaplant@andrew.cmu.edu}

\begin{abstract}
We present predictions for the 21 cm brightness temperature power spectrum during the Epoch of Reionization (EoR). We discuss the implications of the ``light cone'' effect, which incorporates evolution of the neutral hydrogen fraction and 21 cm brightness temperature along the line of sight. Using a novel method calibrated against radiation-hydrodynamic simulations, we model the neutral hydrogen density field and 21 cm signal in large volumes ($L = 2$ Gpc/$h$). The inclusion of the light cone effect leads to a relative decrease of about 50\% in the 21 cm power spectrum on all scales. We also find that the effect is more prominent at the midpoint of reionization and later. The light cone effect also can introduce an anisotropy along the line of sight. By decomposing the 3D power spectrum into components perpendicular to and along the line of sight, we find that in our fiducial reionization model, there is no significant anisotropy. However, parallel modes can contribute up to 40\% more power for shorter reionization scenarios. The scales on which the light cone effect is relevant are comparable to scales where one measures the baryon acoustic oscillation. We argue that due to its large comoving scale and introduction of anisotropy, the light cone effect is important when considering redshift space distortions and future application to the Alcock-Paczy\'{n}ski test for the determination of cosmological parameters.
\end{abstract}


\section{Introduction}
\label{sec:intro}
\addtocounter{footnote}{-1} 
\addtocounter{Hfootnote}{-1}

During the cosmological dark ages, the massive components of the universe were largely cold dark matter and neutral hydrogen. As the first stars and galaxies began to form, the UV photons emitted into the surrounding intergalactic medium (IGM) reionized the hydrogen. This phase transition is known as the Epoch of Reionization (EoR, \citealt{loeb_furlanetto2012}). During the reionization process, it is expected that ionized hydrogen formed bubbles in the IGM surrounding stars, creating patches of reionized gas. As the photons travelled further out into the IGM, the ionized bubbles grew larger, until they eventually joined together. Subsequently, most of the remaining neutral hydrogen was localized to the inside of galaxies, with the rest of the IGM being highly ionized. For reviews of the EoR, see \citet{furlanetto_etal2006}, \citet{morales_wyithe2010}, \citet{loeb_furlanetto2012}, and \citet{pritchard_loeb2012}.

This currently accepted description is overly simplistic because the precise details of reionization are still largely unknown. From observing the Gunn-Peterson absorption trough (\citealt{gunn_peterson1965}) in the Ly$\alpha$ forest, we can infer that the global neutral hydrogen fraction $f_\mathrm{HI}$ was greater than 10$^{-3}$ until $z \sim 6$ (\citealt{fan_etal2006}). Recent probes of the cosmic microwave background radiation (CMB) such as the \textit{Wilkinson Microwave Anisotropy Probe} (\textit{WMAP}) and \textit{Planck} have measured the Thomson optical depth of the IGM, which is a measure of the integrated electron density \citep{hinshaw_etal2012,planck2013}. \textit{WMAP}-9 reports a value of $\tau = 0.089 \pm 0.014$, which assuming an instantaneous reionization gives $z_{\mathrm{reion}} = 10.6 \pm 1.1$. Another experimental constraint comes from using the Hubble Space Telescope Ultra Deep Field observations of the very first galaxies, which contains information about the UV luminosity of star-forming galaxies at early times \citep{robertson_etal2013}.

One of the most promising tools for further probing this epoch comes from the hyperfine transition of neutral hydrogen. The rest-frame wavelength of this transition is $\lambda \approx 21$ cm. The precise nature of the 21 cm signal depends on several factors, including when the midpoint of reionization occurred, the duration of reionization, and the dominant method by which hydrogen is reionized (\textit{e.g.}, ionization via UV vs. x-ray photons). When making a measurement using the 21 cm brightness temperature, one can observe the \textit{global signal} or the \textit{power spectrum}. The former is the brightness temperature average over the entire sky, which during reionization is $\order{10}$ mK. The power spectrum is a statistical measure of the fluctuations in the field as a function of $k$-space. More information about the importance of the 21 cm signal can be found in, for example, \citet{loeb_zaldarriaga2004}, \citet{cooray_2004}, and \citet{bharadwaj_ali2004}.

With the advent of large radio-telescope and dipole arrays constructed specifically to observe the EoR, there have recently been several exciting advances regarding 21 cm observations. Some of the observational probes that are currently taking EoR data (or will be in the near future) are, for example, the Low Frequency Array \citep[LOFAR\footnote{www.lofar.org};][]{harker_etal2010}, the Precision Array for Probing the Epoch of Reionization \citep[PAPER\footnote{eor.berkeley.edu};][]{parsons_etal2010}, the Giant Metrewave Radio Telescope \citep[GMRT\footnote{gmrt.ncra.tifr.res.in};][]{pen_etal2009}, the Murchison Widefield Array \citep[MWA\footnote{www.mwatelescope.org};][]{bowman_etal2005}, and the Experiment to Detect the Global EoR Step \citep[EDGES\footnote{www.haystack.mit.edu/ast/arrays/Edges};][]{bowman_rogers2010}. These arrays are designed to extract the 21 cm signal over a relatively narrow frequency band, targeting a particular redshift. An upcoming telescope, such as the Square Kilometer Array \citep[SKA\footnote{www.skatelescope.org};][]{mellema_etal2013}, will be designed to take full tomographic data of the EoR, and map the 21 cm signal as a function of frequency.

When performing a three dimensional measurement of the 21 cm signal, there are several important caveats to bear in mind. Two of the major effects are the light cone effect and redshift space distortions (RSD). The light cone effect comes purely from the time delay of propagation of the signal to the observer. In general, different comoving distances from an observer correspond to different points in redshift space. For sufficiently large scales, the comoving distance spanned by the observed volume corresponds to a large duration in redshift space. The neutral hydrogen fraction can change significantly if the length of the observed redshift interval is comparable to or larger than the duration of reionization. This evolution of the neutral fraction also introduces anisotropy along the line of sight in the 3D power spectrum. The light cone effect has been explored with respect to 21 cm observations semi-analytically by \citet{barkana_loeb2004} and numerically by \citet{datta_etal2012}. In previous works, the light cone was deemed to have an $\order{1}$ effect on the 21 cm brightness temperature two-point correlation function or power spectrum, respectively. We show in this work that the light cone can have a similar effect for sufficiently large volumes. Furthermore, we show that the light cone is most important around the midpoint of reionization, where $0.4 \lesssim f_\mathrm{HI} \lesssim 0.6$.

RSD are the result of peculiar velocities of the signal sources. Since the simplest computation of the 21 cm signal assumes that the only source of velocity is the Hubble flow, peculiar velocities lead to a correction of the predicted signal. The effect of RSD has already been applied to 21 cm cosmology \citep[\textit{e.g.},][]{barkana_loeb2005,bharadwaj_ali2005,mao_etal2012,jensen_etal2013,shapiro_etal2013,majumdar_etal2013}. In general, RSD have an $\order{1}$ effect on the 3D 21 cm brightness temperature power spectrum at the largest scales. The effects of RSD are thought to be most prominent early in reionization. For example, \citet{jensen_etal2013} show that RSD are most important for $0.7 \lesssim f_\mathrm{HI} \lesssim 1.0$, peak at $f_\mathrm{HI} \sim 0.9$, and have little impact after the midpoint of reionization.

The light cone effect also has important implications for measurements that use the baryon acoustic oscillation (BAO) method. The BAO method is important for understanding the accelerating expansion of the universe, and is used to make measurements of fundamental parameters such as $H(z)$. The BAO scale is large, typically 150 comoving Mpc. As we show, the light cone effect also becomes important on these scales. The BAO method can be subjected to the Alcock-Paczy\'{n}ski test \citep{alcock_paczynski1979}, which uses spherical features and relates their angular diameter distance to their extent in redshift space to determine cosmological parameters. Proper application of this test requires an accurate understanding of any anisotropies between perpendicular and parallel behavior of these features. As is discussed more in the body of this paper, the light cone effect can introduce anisotropy in the 21 cm signal in the parallel direction. Therefore, if the 21 cm signal is to be used in BAO methods, the light cone effect must be properly understood and included in calculations. For application of the BAO method to the 21 cm signal, see \citet{nusser2005} and \citet{barkana2006}; for discussion of the BAO theory and current implementations, see \citet{weinberg_etal2012}.

Our approach combines numerical simulations with semi-analytic tools. We first perform a reionization simulation including hydrodynamics and radiative transfer on a relatively small volume. Once a statistical measure has been devised for how the matter overdensity field is related to the redshift of reionization, this statistical measure is used on a matter-only simulation in a larger volume that still accurately predicts reionization observables. In addition, different reionization histories can be explored rapidly without rerunning computationally expensive simulations. For a more thorough explanation of the general method outlined here, see \citet{battaglia_etal2012a}. For applications of this method to EoR observables related to the CMB, see \citet{natarajan_etal2012} and \citet{battaglia_etal2012b}.

The main purpose of this paper is to quantify how the 21 cm power spectrum signal changes with the inclusion of the light cone effect. In \S\ref{sec:methodology}, we discuss the methodology behind the analysis and briefly describe the numerical techniques being applied. In \S\ref{sec:analysis}, we discuss the basic science of the 21 cm brightness temperature power spectrum, and the types of statistical tests we perform on the data. Also in this section, we examine the application of these tests to data which comes from performing the analysis on a simulation box at a single redshift snapshot. (Hereafter, we refer to this type of data as ``coeval cubes.'') In \S\ref{sec:lc}, we discuss the light cone effect on the 3D power spectrum. Then, in \S\ref{sec:observation}, we talk about specific applications to various observational endeavors, and how this signal might appear in real-world measurements. In \S\ref{sec:discussion}, we discuss other effects and potential difficulties related to the 21 cm signal. To conclude, in \S\ref{sec:conclusion}, we talk about future prospects and outlooks. We assume a $\Lambda$CDM cosmology with $\Omega_{\Lambda} = 0.73$, $\Omega_m = 0.27$, $\Omega_b = 0.045$, $h = 0.70$, and $\sigma_8 = 0.80$. These values are consistent with the \textit{WMAP}-9 results \citep{hinshaw_etal2012}.

\section{Methodology}
\label{sec:methodology}
In Paper I \citep{battaglia_etal2012a} we developed a semi-analytic model for relating the matter content in a computational simulation cell with the redshift at which the cell becomes 90\% ionized. This approach exploits the fact that the matter overdensity field, defined as
\begin{equation}
\delta_m(\va{x}) \equiv \frac{\rho_m(\va{x}) - \bar{\rho}_m}{\bar{\rho}_m},
\label{eqn:deltam}
\end{equation}
is highly correlated with fluctuations in the redshift of reionization field ($z_{\mathrm{re}}(\va{x})$) defined as
\begin{equation}
\delta_z(\va{x}) \equiv \frac{[z_\mathrm{re}(\va{x}) + 1] - [\bar{z}+1]}{\bar{z}+1},
\label{eqn:deltaz}
\end{equation}
on large scales ($\gtrsim 1$ Mpc/$h$) \citep{battaglia_etal2012a}. To motivate this observation, note that in an ``inside-out'' reionization scenario, the densest regions are the ones which form stars and galaxies capable of producing reionizing photons the earliest. The difference in amplitude between the two fields can be quantified using the bias parameter $b_{zm}(k)$ which is applied to the two fields in Fourier space. The bias parameter can be written as:
\begin{equation}
b_{zm}^2 (k) \equiv \frac{\ev{\delta_z^* \delta_z}_k}{\ev{\delta_m^* \delta_m}_k} =\frac{P_{zz}(k)}{P_{mm}(k)},
\label{eqn:psbias}
\end{equation}
where $P_{xx}(k)$ is the auto-power spectrum of a field $\delta_x$. In order to quantify how similar two fields are, the cross-correlation coefficient $r$ can be used. This quantity can be defined as:
\begin{equation}
r_{zm} (k) \equiv \frac{\langle \delta_z^* \delta_m \rangle_k}{\sqrt{ \langle \delta_z^2 \rangle_k \langle \delta_m^2 \rangle_k}} = \frac{P_{zm}(k)}{\sqrt{P_{zz}(k) P_{mm}(k)}},
\label{eqn:cc}
\end{equation}
where $P_{xy}(k)$ is the 3D cross-power spectrum of the fields $\delta_x$ and $\delta_y$. The normalization ensures that $r \in [-1,1]$. For values where the cross-correlation coefficient becomes 1, the fields are highly correlated, and the amplitudes of the fields differ only by their bias factor. This is true for the matter and reionization fields during the EoR on large scales \citep{battaglia_etal2012a}.

Since the matter and reionization fields are highly correlated on large scales, the bias parameter can be used to relate their amplitude difference. In general, the bias will change as a function of $k$. We have chosen a functional form of the bias defined in Equation~(\ref{eqn:psbias}) in such a way to reproduce the relationship observed in simulations. We define this bias $b_{zm}$ to be:
\begin{equation}
b_{zm} = \frac{b_0}{\qty(1 + \frac{k}{k_0})^\alpha}.
\label{eqn:bias}
\end{equation}
There are essentially three free parameters in this model: $b_0$, $k_0$, and $\alpha$. The value of $b_0$ can be predicted using excursion set formalism in the limit that $k \to 0$ \citep{barkana_loeb2004}. We have chosen $b_0$ to be 0.593.

In order to determine best-fit values for the parameters $k_0$ and $\alpha$, we compare the matter overdensity and reionization-redshift fields using a RadHydro code, which contains radiative transfer + hydrodynamics + $N$-body simulation \citep{trac_etal2008}. These particular simulations contain $2048^3$ dark matter particles, $2048^3$ gas cells, and 17 billion adaptive rays in a 100 Mpc/$h$ cubical box. We find that the best fits for the values were $\alpha=0.564$ and $k_0=0.185$ $h$ Mpc$^{-1}$. In addition to these physically motivated ``fiducial'' values, two other sets of values were chosen to represent more extreme reionization scenarios: a long and short reionization history, parameterized in our model with the values of $(\alpha, k_0)$ = $\{(1.8, 0.1)$, $(0.2, 0.9)\}$, respectively. Examining different reionization histories allows for the identification of features in the power spectrum which may indicate how quickly reionization occurred.

Once the values in the bias relationship have been fixed, the matter overdensity field can be used in order to construct the reionization-redshift field. Accordingly, we performed a dark-matter-only simulation with a particle-particle-particle-mesh ($\mathrm{P^3M}$) $N$-body code using $2048^3$ dark matter particles in a 2 Gpc/$h$ box. Then, using a snapshot of the matter overdensity field at the midpoint of reionization $\bar{z}$, we apply the bias relation in Equation~(\ref{eqn:bias}). For more details on this method, see \citet{battaglia_etal2012a}.

\begin{figure*}[t]
  \centering
  \resizebox{0.5\hsize}{!}{\includegraphics{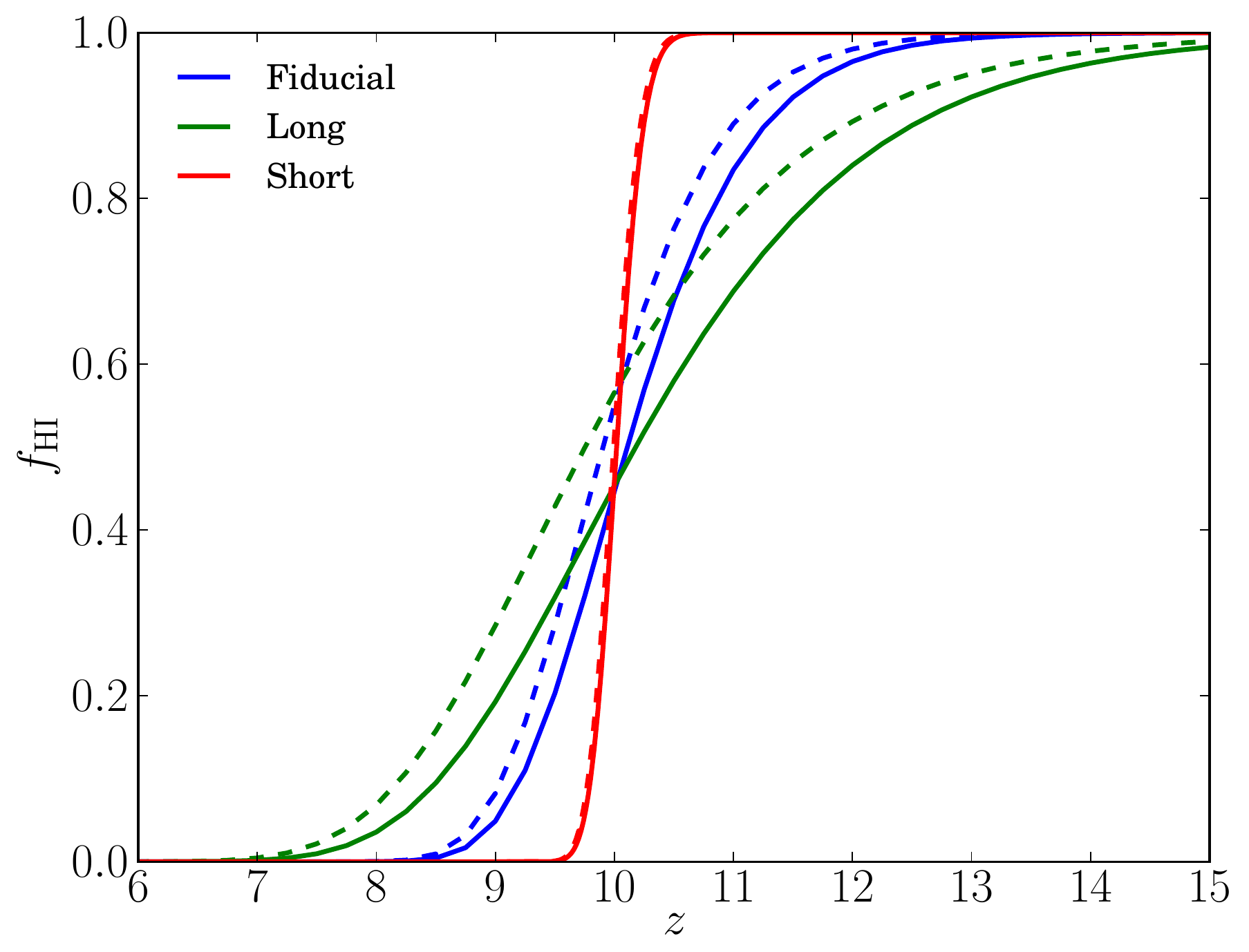}}%
  \resizebox{0.5\hsize}{!}{\includegraphics{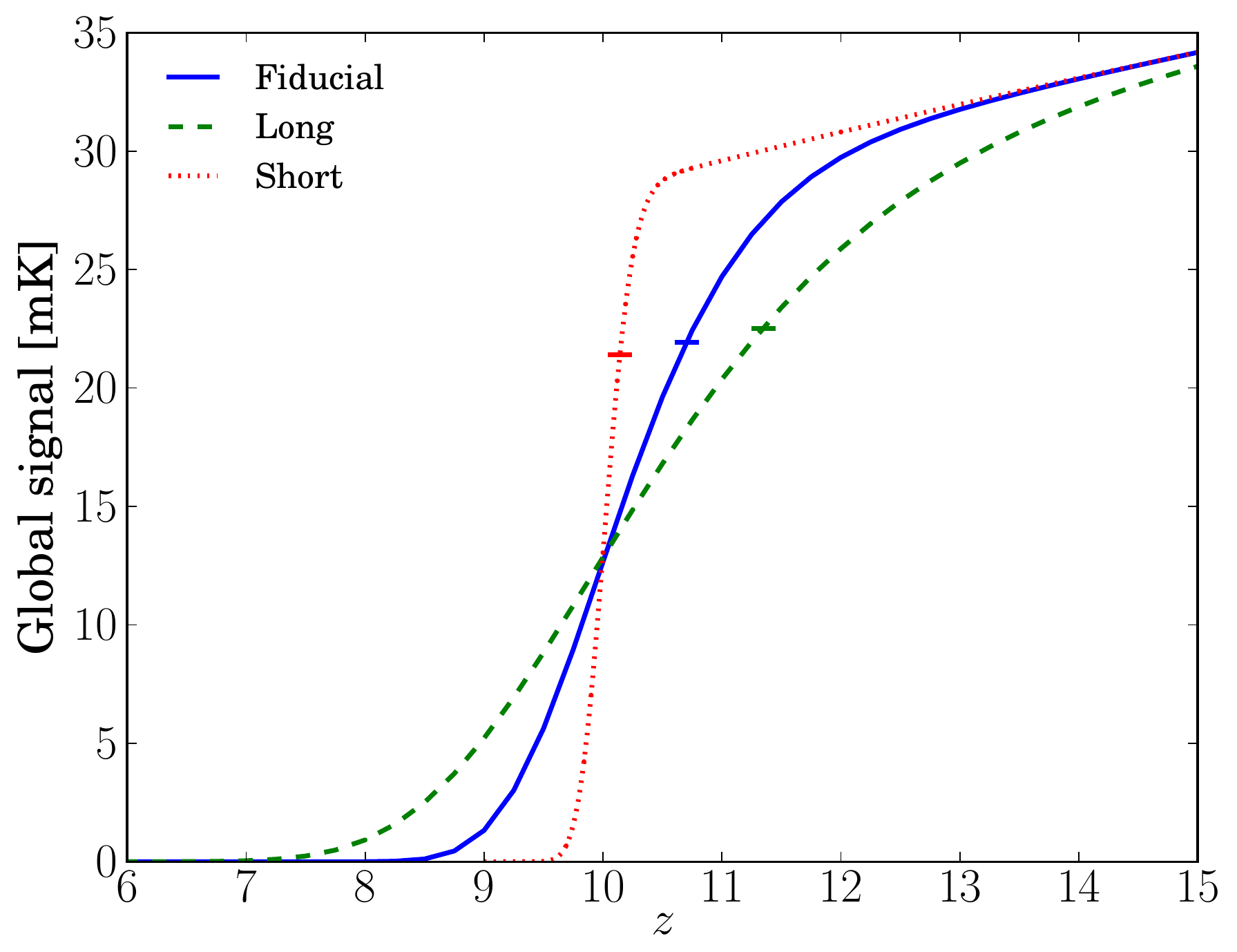}}\\
  \caption{Left: the average neutral hydrogen fraction as a function of redshift. Plotted are the mass-weighted average (solid lines) and the volume-weighted average (dashed lines). In an inside-out reionization scenario, the densest regions of the universe are the first ones to reionize, so the mass-weighted neutral fraction is always lower than the volume-weighted one. Right: the global 21 cm signal as a function of redshift for the different reionization histories from Equation~(\ref{eqn:tb}), with the simplifying assumption that $T_S \gg T_\mathrm{CMB}$. This approximation is only physically justifiable for mass-weighted neutral fractions of $f_\mathrm{HI} \lesssim 0.75$; nevertheless, we plotted the global temperature predicted by Equation~(\ref{eqn:tb}) for higher redshifts since it is still approximately true in this redshift range. We have marked the points where $f_\mathrm{HI} \sim 0.75$ by small ticks on the lines. By construction, all of the histories have the same midpoint of reionization of $\bar{z}=10$, which accounts for the point of intersection.}
  \label{fig:global}
\end{figure*}

Figure~\ref{fig:global} shows a plot of the ionization fraction of the simulation volume, both mass- and volume-weighted. All neutral fractions reported in the rest of this paper, unless otherwise noted, are mass-weighted. The duration of reionization is measured by finding the redshift range for when the simulation cube is 25\% ionized to 75\% ionized, which measures the ``50\% ionization width'' $\Delta z_{50}$.  The 50\% reionization duration in redshift for the long, fiducial, and short cases (weighted by mass) are: $\Delta z_{50} = 2.11$, 1.10, and 0.24. The reionization model considered here does not allow for ``exotic'' reionization scenarios, such as extended reionization or recombination before a second ionization.

\section{Analysis}
\label{sec:analysis}

\subsection{21 cm Theory}
\label{sec:21cm}
The 21 cm signal tracks regions of neutral hydrogen in the IGM. The application of the radiative transfer equation to CMB photons free-streaming from the surface of last scattering and passing through neutral hydrogen in the intergalactic medium predicts whether the neutral hydrogen will absorb or emit radiation at 21 cm.

The difference between the brightness temperature and the temperature of the CMB is given as \citep{madau_etal1997,harker_etal2010}:
\begin{align}
\frac{\delta T_b}{\mathrm{mK}} &= 38.6 \, h (1 + \delta_m) x_{\mathrm{HI}} \qty(\frac{T_S - T_\mathrm{CMB}}{T_S}) \notag \\
& \qquad \times \qty( \frac{\Omega_b}{0.045} ) \qty[ \qty( \frac{0.27}{\Omega_m} ) \qty( \frac{1 + z}{10} ) ]^{\frac{1}{2}} \label{eqn:tb} \\[0.5em]
&= T_0 (z) (1 + \delta_m) x_{\mathrm{HI}}, \notag
\end{align}
where $x_{\mathrm{HI}}$ is the neutral hydrogen fraction (assumed to be 0 or 1 for an individual gas cell), and $T_0$ is the redshift-dependent ``average temperature'' of the signal, which is modulated by the spatial fluctuations of the matter overdensity field and the ionization state. This analysis was performed in a regime where $\Omega_\Lambda$ can be safely ignored. Equation~\ref{eqn:tb} gives the difference of the brightness temperature at a frequency corresponding to 21 cm from the CMB as a function of redshift and spatial position.

In the following analysis, it has also been assumed that the spin temperature is large compared to the CMB temperature, $T_S \gg T_\mathrm{CMB}$. Following from the results of \citet{santos_etal2008}, this factor is approximately 1 for mass-weighted neutral fractions $f_{\mathrm{HI}} \lesssim 0.75$. Once the neutral fraction has reached this value, the spin temperature is collisionally coupled to the kinetic temperature of the gas, which is typically 2 orders of magnitude larger than the effective CMB temperature. This assumption is applicable to a large range of reionization scenarios, \textit{e.g.}, ones where UV photons from stars photo-ionize and photo-heat the neutral hydrogen, so that hydrogen's spin temperature couples to the kinetic energy of the gas particles and becomes much hotter than the CMB. Exotic reionization scenarios, \textit{e.g.}, those where reionization is caused by x-ray heating, do not necessarily meet the condition that $T_S \gg T_\mathrm{CMB}$. However, these scenarios have not been examined in this analysis, and these considerations have been saved for future work.

Figure~\ref{fig:global} shows the global 21 cm signal as a function of redshift for the different reionization scenarios. The duration of the reionization history affects the rate at which the global signal diminishes: the long reionization scenario drops gradually, whereas the short reionization scenario drop rapidly. Observationally, the signal from a shorter reionization scenario is easier to measure than a longer one \citep{bowman_rogers2010}, since a shorter reionization scenario would appear as a sharper feature in frequency space.

\subsection{3D Power Spectrum}
\label{sec:ps}
We define the 3D power spectrum as $P_{xx}(k) = \ev{\delta_x^* \delta_x}_k$, and the dimensionless power spectrum $\Delta^2(k)~\equiv~k^3 P(k)/2\pi^2$. The features of the coeval matter overdensity field power spectrum have already been extensively explored, so we will only list some common features. As the universe evolves over time, the amplitude of the matter power spectrum increases monotonically. Because the 21 cm brightness temperature (\textit{cf.} Equation~(\ref{eqn:tb})) is proportional to the matter overdensity field, one might expect the 21 cm brightness temperature power spectrum also to increase monotonically. However, the 21 cm signal also incorporates the neutral hydrogen fraction, and so as the universe becomes increasingly ionized, the signal diminishes. This evolution causes the amplitude of the 21 cm signal to increase as the universe begins to ionize, peak at a particular neutral fraction, and then decrease as the universe ionizes further. The shape of the 21 cm power spectrum in the coeval case has also been examined \citep[\textit{e.g.,}][]{lidz_etal2008}, and the value corresponding to a peak in large scale power is $\sim$50\% ionization fraction.

\begin{figure*}[t]
  \centering
  \resizebox{0.49\hsize}{!}{\includegraphics{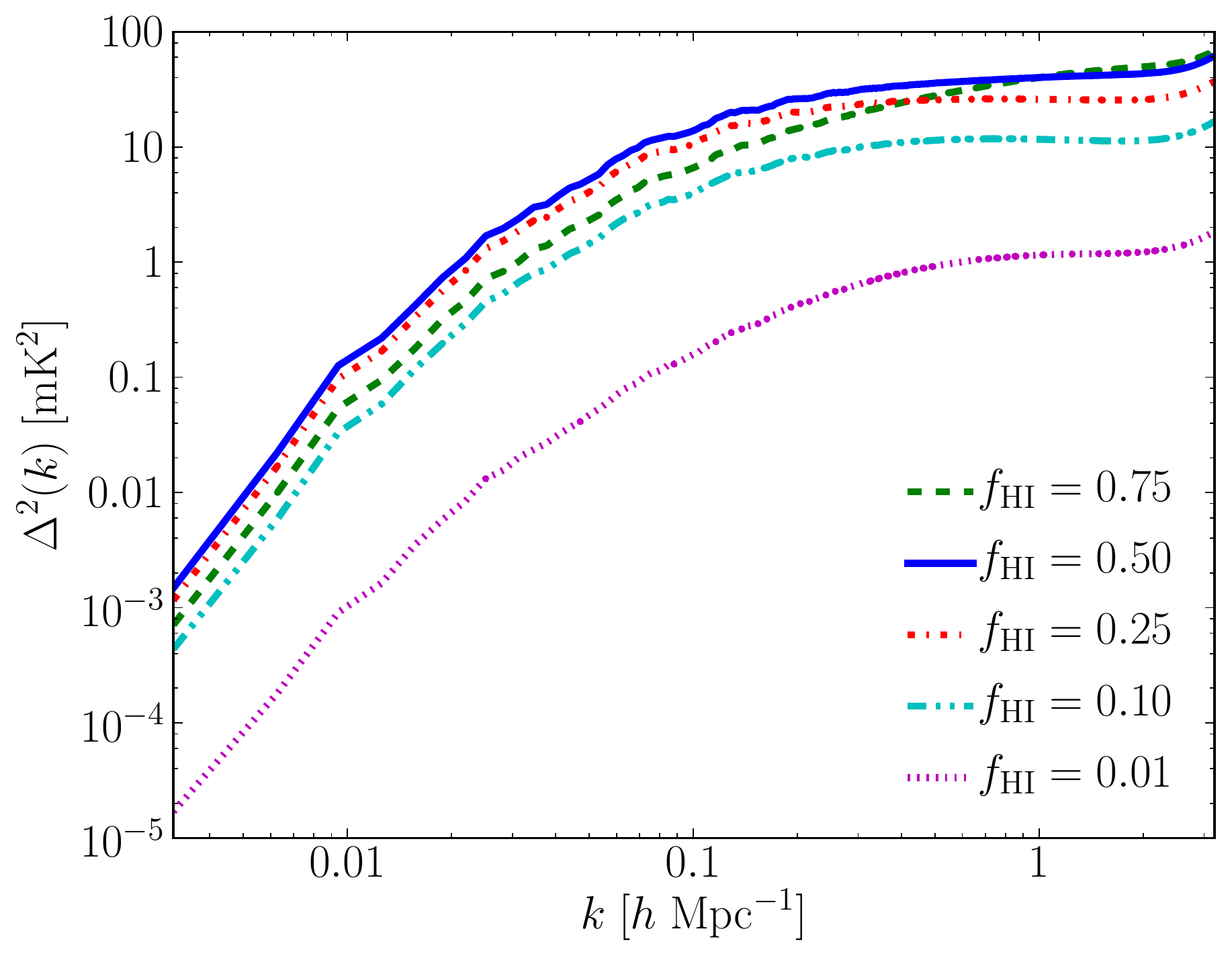}}%
  \hspace{2pt}%
  \resizebox{0.49\hsize}{!}{\includegraphics{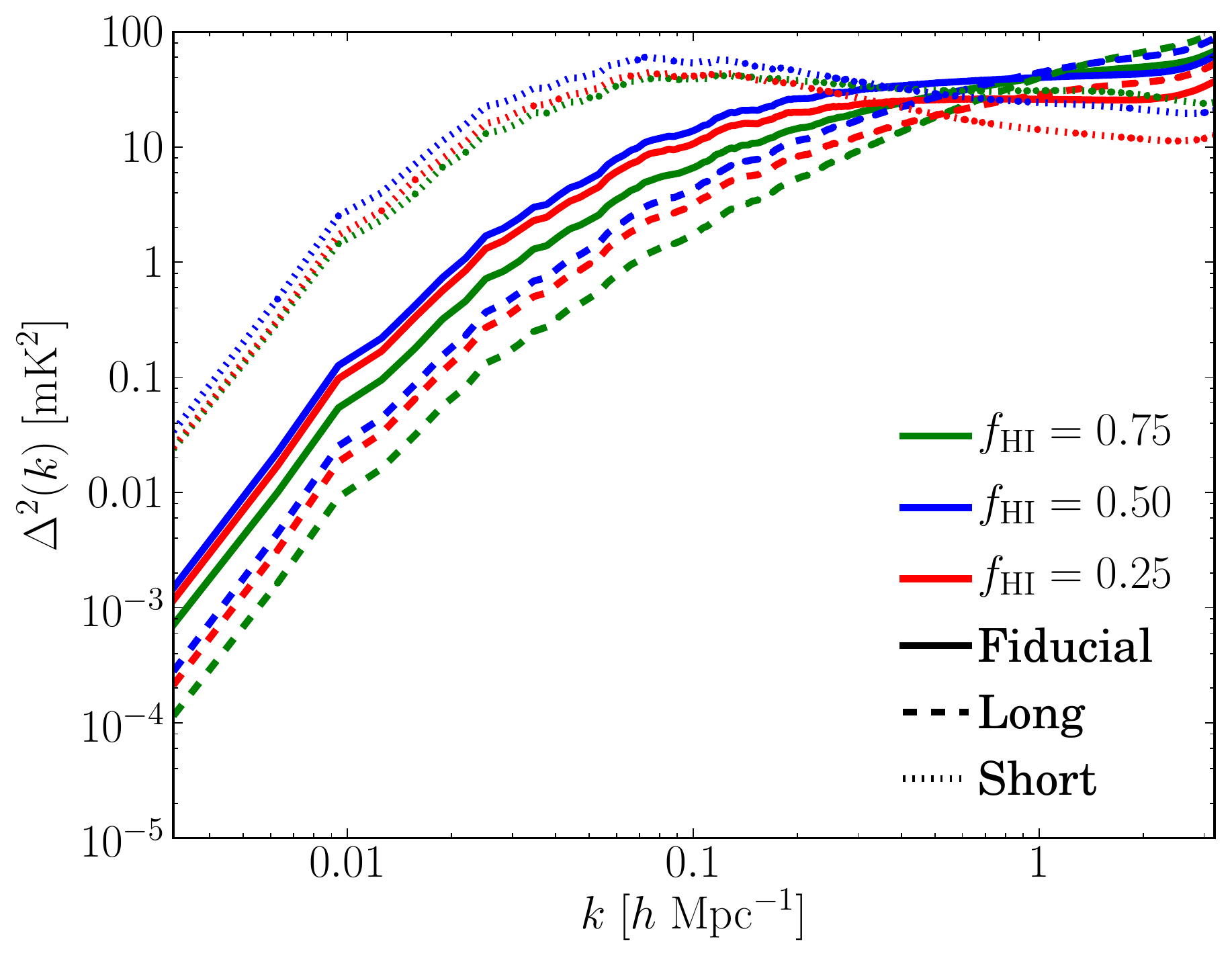}}\\
  \caption{Left: a plot of the 3D 21 cm brightness temperature power spectrum, as a function of neutral hydrogen fraction. On large scales, the power peaks at $f_{\mathrm{HI}} \sim 0.5$, but at smaller scales it peaks for a larger neutral fraction, $f_{\mathrm{HI}} \sim 0.75$. The reason for this is that at a larger neutral fraction, only the densest regions are ionized, so the 21 cm power spectrum looks more like the matter power spectrum. At a neutral fraction of less than 50\%, the differences in amplitude for different values of $f_\mathrm{HI}$ on large scales is roughly proportional to the difference in neutral fraction. This phenomenon is due to the fact that the redshift evolution of the 21 cm signal after the midpoint is dominated by the changing neutral fraction. Right: the evolution of the power spectrum for different reionization histories. Across all reionization histories, the power spectrum is larger near 50\% ionization. The shape of the power spectrum changes dramatically for different reionization histories, where in general, a shorter duration of reionization implies more large-scale power and less small-scale power.}
  \label{fig:3d-fid-ps}
\end{figure*}

We calculate the power spectrum as a function of neutral fraction, since equal neutral fractions between reionization scenarios capture the same physics better than equal redshifts. We linearly interpolate between matter overdensity fields from adjacent snapshots in order to create a power spectrum as a function of specific neutral fractions. The 21 cm brightness field was computed from this interpolated matter overdensity field using Equation~(\ref{eqn:tb}) where $x_\mathrm{HI}$ was determined from $z_\mathrm{re}(\va{x})$.

Figure~\ref{fig:3d-fid-ps} shows the features of the 3D power spectrum from the fiducial reionization scenario. On large scales, the amplitude peaks near the midpoint of reionization, $f_\mathrm{HI} \sim 0.5$. On small scales, the power is largest for the highest neutral fraction, $f_\mathrm{HI} \sim 0.75$. Early on in reionization, only the densest regions have become ionized, which means the 21 cm brightness temperature power spectrum looks very similar to the matter power spectrum. As the universe becomes more ionized, this small-scale power is lost due to the ionized regions growing larger.

Figure~\ref{fig:3d-fid-ps} also shows the 3D power spectrum across the different reionization scenarios of our model. The general shape of the spectra changes dramatically as a function of reionization history: as the duration of reionization decreases, more power is transferred from small scales to large ones. For our model, although the underlying matter overdensity field is identical across the simulations, the reionization history dramatically changes the predicted shape of the 21 cm power spectrum.

\subsection{Bias Parameter and Average Bias}
\label{sec:cc}

\begin{figure*}[t]
  \centering
  \resizebox{0.47\hsize}{!}{\includegraphics{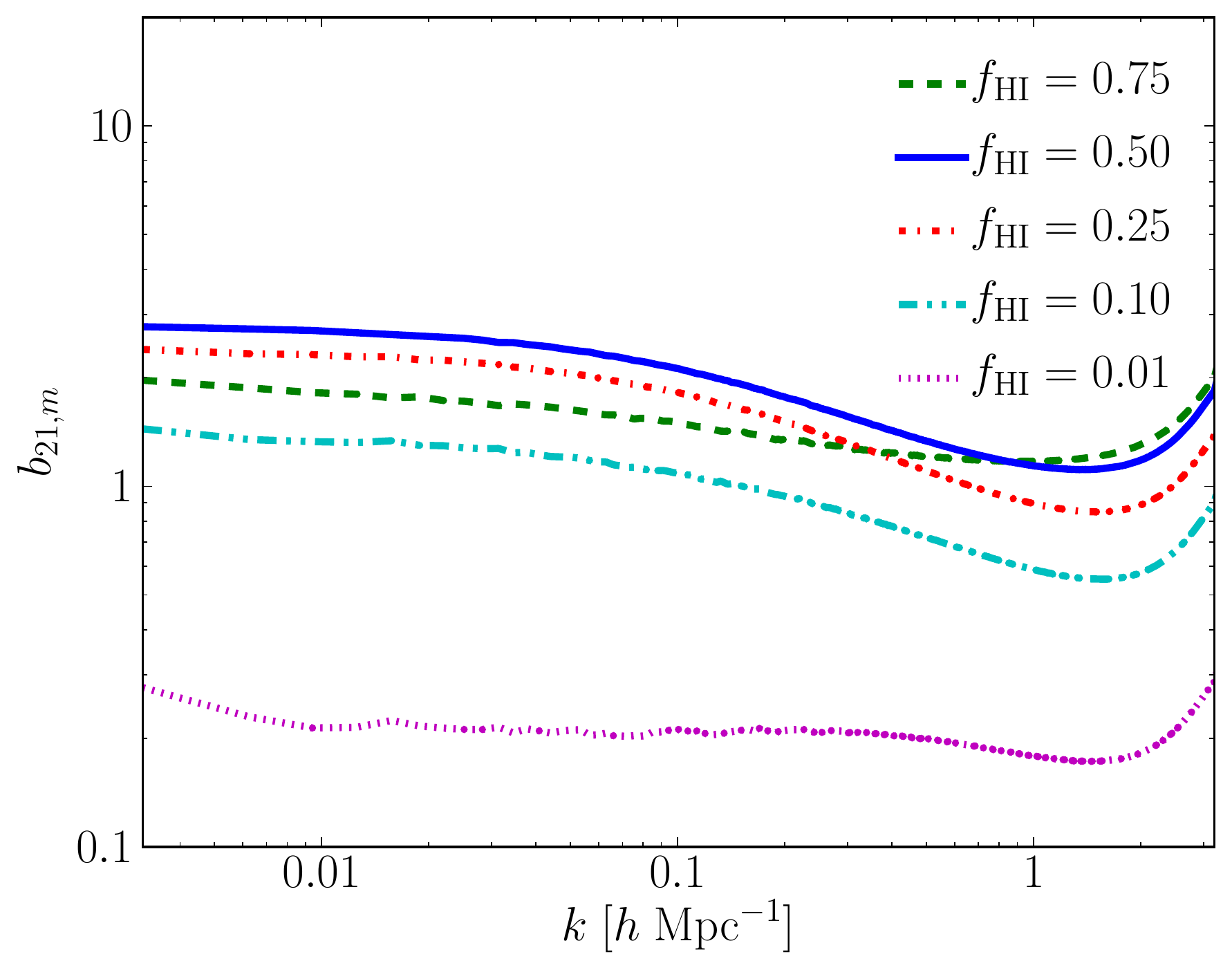}}%
  \hspace{2pt}%
  \resizebox{0.48\hsize}{!}{\includegraphics{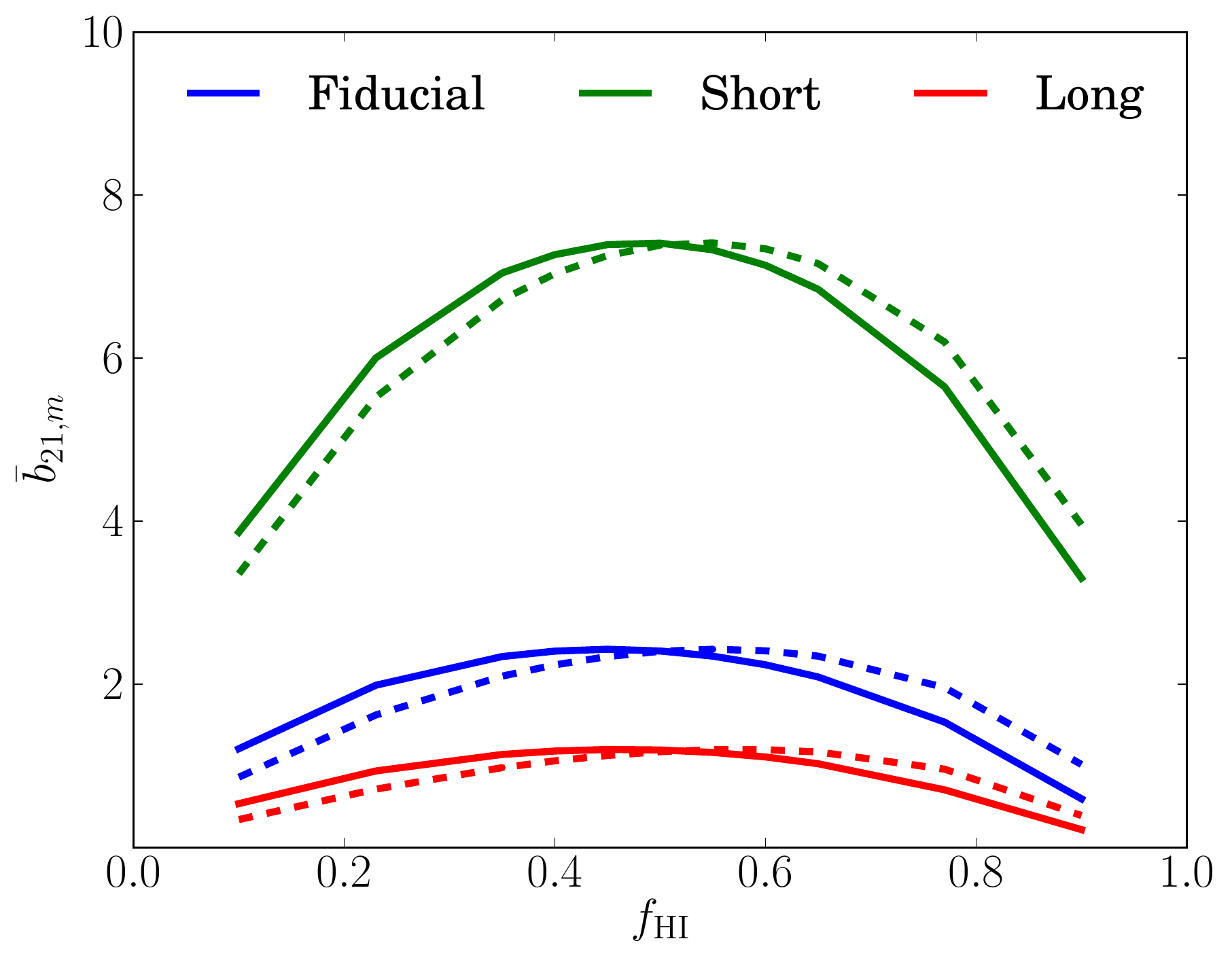}}\\
  \caption{Left: the scale-dependent bias between the two fields, defined in Equation~(\ref{eqn:psbias}), for the fiducial reionization history. To remove the redshift dependence between different neutral fractions, we divide the 21 cm brightness temperature by $T_0(z)$ defined in Equation~(\ref{eqn:tb}). One can see that the value is fairly constant in the region $k \lesssim 0.1$ $h$ Mpc$^{-1}$, leading to the choosing of this value for the large-scale bias parameter. The small-scale structure for large $k$-values changes noticeably as the universe becomes more ionized. Right: A plot of the large scale bias relationship between the 21 cm power spectrum and matter power spectrum at different neutral fractions. Shown are the mass-weighted neutral fraction (solid lines) and the volume-weighted neutral fraction (dashed lines). The bias is calculated according to Equation~(\ref{eqn:avg-bias}), which only takes into account the largest scales ($k < 0.1 \, h\, \mathrm{Mpc^{-1}}$). When the bias is largest, there is the most 21 cm signal relative to the underlying matter overdensity field. Note that as the reionization history becomes shorter, the bias becomes larger at all neutral fractions.}
  \label{fig:bias}
\end{figure*}

Figure~\ref{fig:bias} shows a plot of the bias parameter (Equation~(\ref{eqn:psbias})) between the 21 cm brightness temperature field and matter overdensity field. (The cross-correlation coefficient is discussed further in \S\ref{sec:cclc}.) As already mentioned, the bias parameter can be used to quantify the relative amplitudes between the different power spectra. This quantity has already been applied in a number of settings (\textit{e.g.}, \citealt{fry_gaztanaga1993}, \citealt{heavens_etal1998}, \citealt{croft_etal2002}, etc.). The application at hand is the bias factor between the matter overdensity field and the 21 cm brightness temperature field. Note that for the calculation of the bias parameter, the average temperature $T_0(z)$ is divided out in Equation~(\ref{eqn:tb}) (resulting in $\widehat{\delta T}_b = (1 + \delta_m) x_{\mathrm{HI}}$), in order to remove dependence on redshift. For high values of the neutral fraction, the bias is flatter, meaning that the 21 cm brightness temperature field is more similar to the matter overdensity field. As the universe becomes more ionized, the bias changes more dramatically as a function of $k$. As in the case of the 3D power spectrum, the amplitude of the bias on large scales peaks at $f_\mathrm{HI} \sim 0.5$. The evolution of the matter overdensity field is small compared to the change in the 21 cm brightness temperature.

In regions where the bias is roughly constant, an ``average bias'' can be defined as:
\begin{equation}
\bar{b}_{21,m} = \ev{\sqrt{\frac{\delta_{21}^* \delta_{21}}{\delta_m^* \delta_m}}}_{k<k_\star}
\label{eqn:avg-bias}
\end{equation}
where $k_\star$ (0.1 $h$ Mpc$^{-1}$) is a predefined cutoff value to ensure that the selected regime is relatively constant. For a given reionization history, the power spectrum of both the 21 cm field and the matter overdensity fields is calculated. The average of the ratio of the two power spectra is computed for all $k$-values out to $k_\star$ at several different values of the neutral fraction. The large scale bias is important because it predicts the amplitude of the 21 cm power spectrum compared to the matter power spectrum, especially at large scales. As in the case of the scale-dependent bias, the average temperature of the 21 cm brightness temperature field $T_0(z)$ has beed divided out.

Figure~\ref{fig:bias} shows the average bias for the three different reionization scenarios. As seen in the figure, the bias peaks at an ionization fraction of roughly 50\% by mass. As already discussed in \S\ref{sec:ps}, this coincides with the peak in the power of the 21 cm power spectrum. A large value of the bias implies that the sources of reionization are themselves ``highly biased,'' in the sense that they are larger and rarer for larger values of the bias parameter. Figure~\ref{fig:lc-fields}, which shows the 21 cm brightness field in the coeval and light cone cases, demonstrates this visually. In the coeval column on the left, the short reionization scenario has larger but fewer ionized regions, which implies that the sources are massive and rare. Indeed, the difference between the large voids in the case of short reionization and the small pockets of ionized gas in the long reionization is striking. Thus, the large scale bias parameter is important not only because it yields valuable information about the relation between the 21 cm brightness temperature and the matter overdensity field, but also because it is related to the sources of reionization.

\section{Light Cone Effect}
\label{sec:lc}
The light cone effect on 21 cm power spectra has been examined semi-analytically in \cite{barkana_loeb2006} and numerically in \cite{datta_etal2012}. The previous numerical work was concerned only with relatively small volumes, and found that the light cone effect is an $\order{1}$ effect on their largest scales. We examined the impact of the light cone effect on volumes larger than those used by \citet{datta_etal2012}, and we conclude that this effect is an essential consideration for 21 cm measurements.

\begin{figure*}[p]
  \centering
  Long \\ \includegraphics[width=0.9\textwidth]{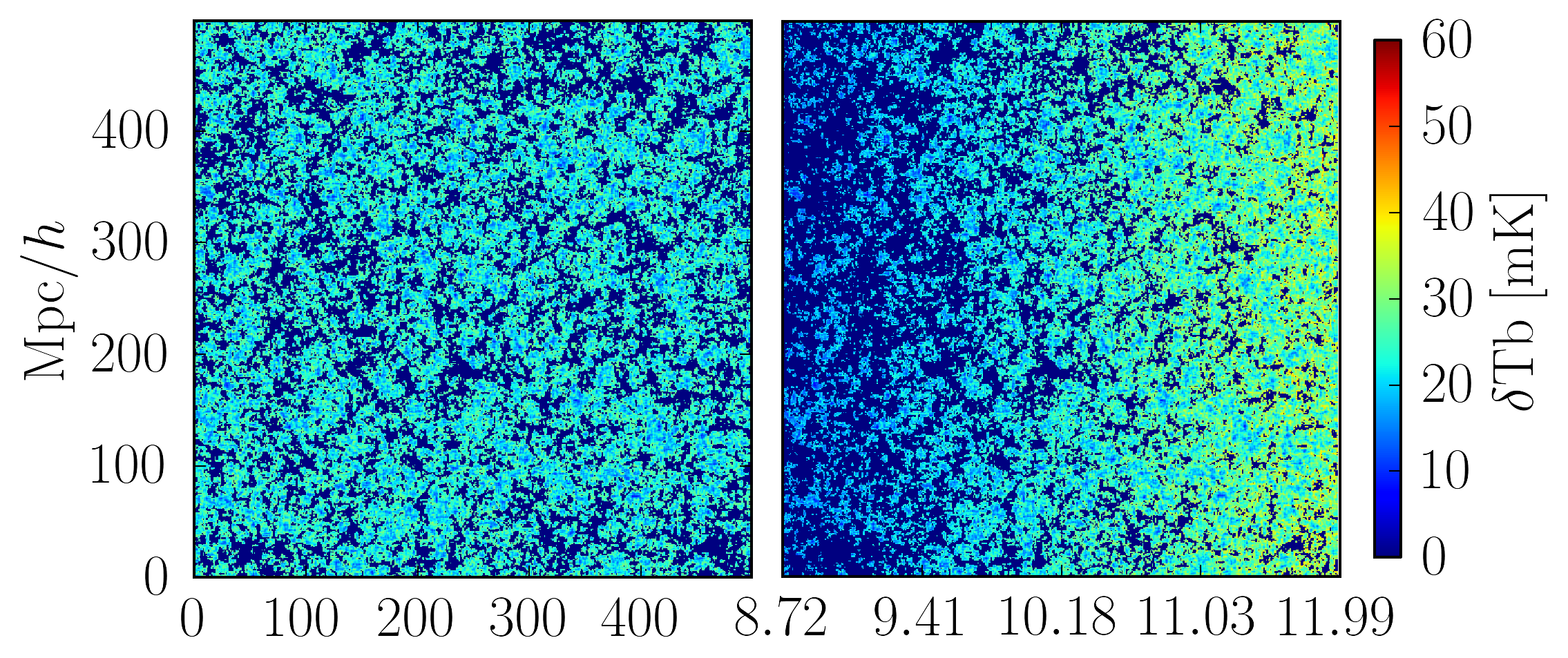}\\
  Fiducial \\ \includegraphics[width=0.9\textwidth]{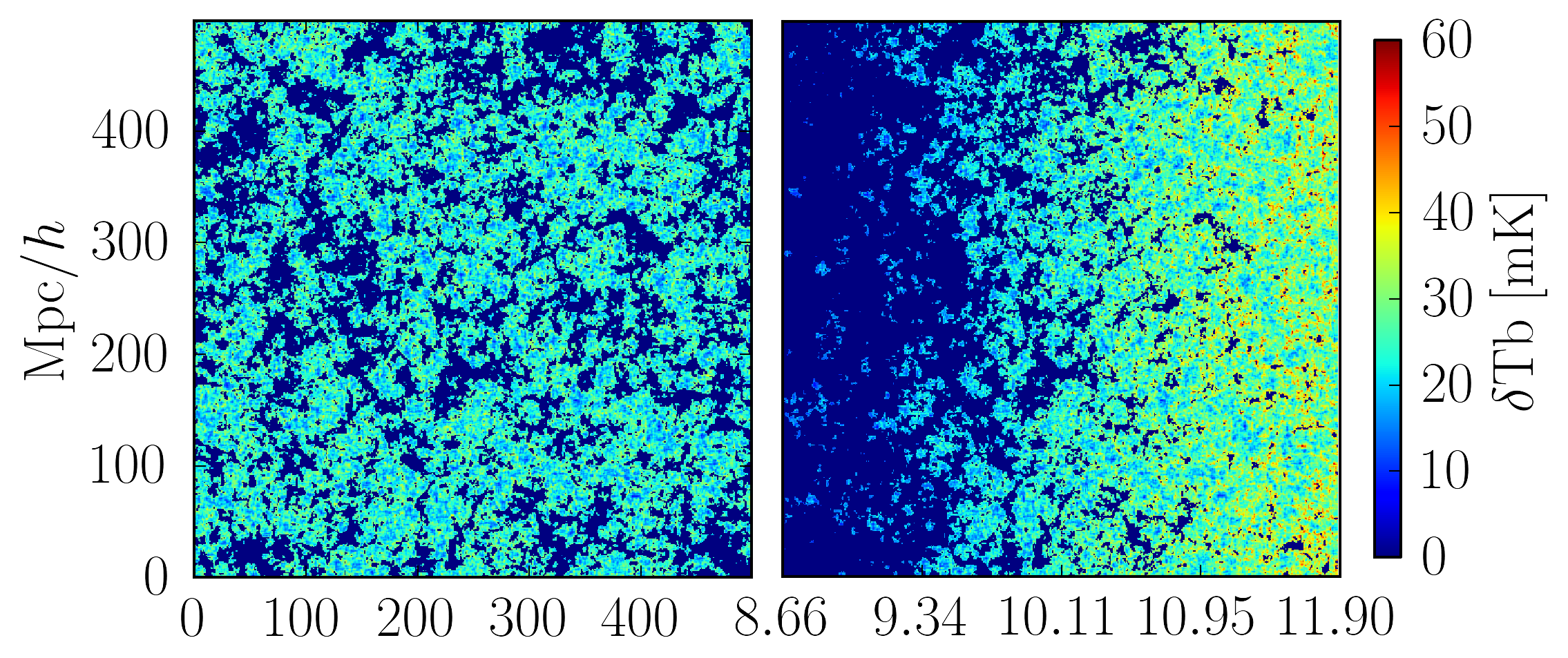}\\
  Short \\ \includegraphics[width=0.9\textwidth]{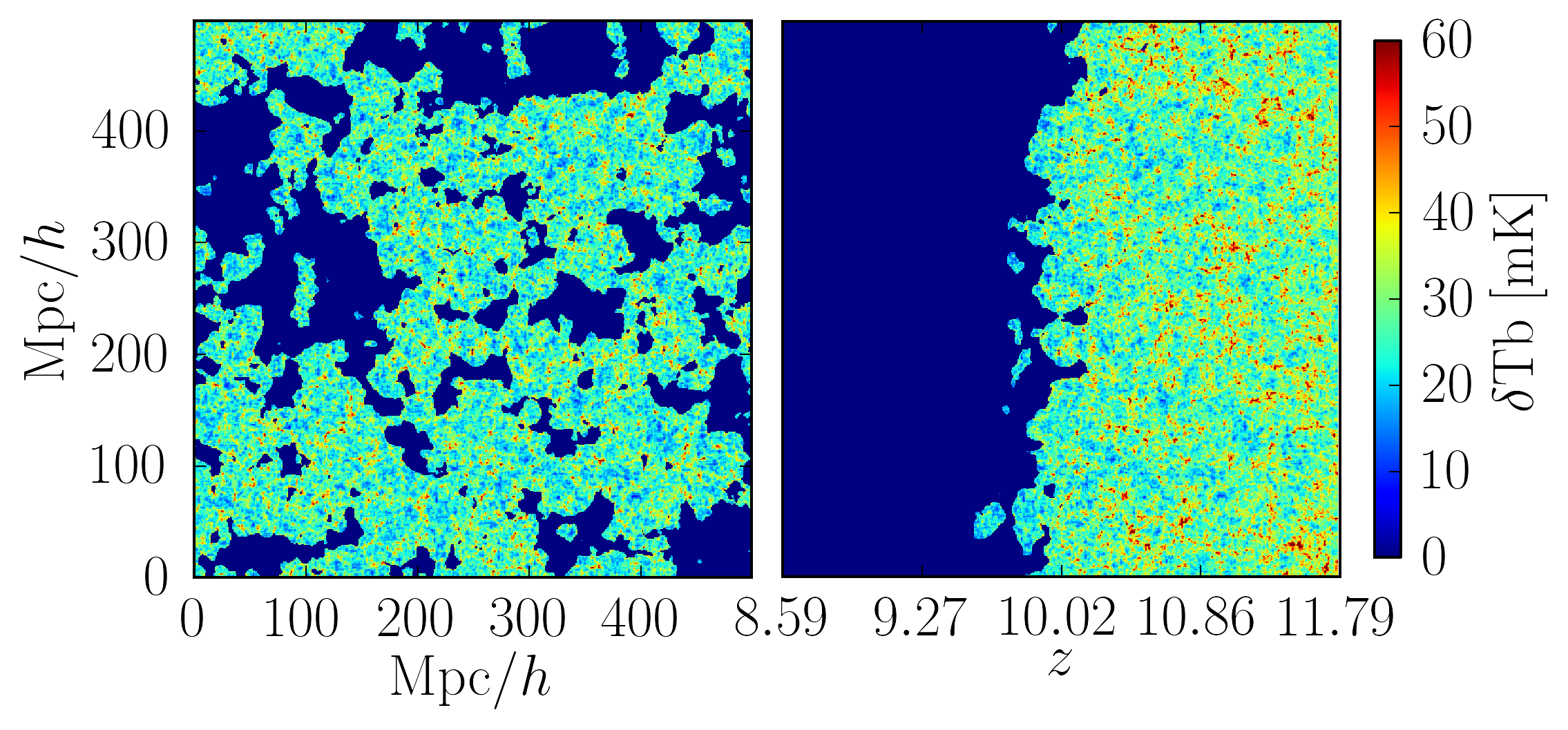}\\
  \caption{A visualization of the evolution of the 21 cm brightness of the simulation cube. Left: a coeval sub-box at 50\% ionization fraction with side length of 500 Mpc/$h$ for long (top), fiducial (middle), and short (bottom) reionization scenarios. Right: the corresponding light cone cube, which includes evolution of the ionization field. The $x$-axis on the right shows the redshift instead of the comoving distance, with the center of the box placed at the redshift equalling 50\% ionization by mass.  We notice that the 21 cm signal initially follows the underlying matter fluctuations at the side of the box farther from the observer where it is almost entirely neutral, then gradually fades to zero brightness as the IGM becomes increasingly ionized. For the long reionization scenario, the coeval case has smaller bubble sizes at 50\% reionization, and the light cone effect is not as pronounced. For the short reionization scenario, the coeval case has very pronounced bubbles of ionized gas at 50\% reionization, and the light cone effect is quite dramatic.}
  \label{fig:lc-fields}
\end{figure*}

In essence, the light cone effect is due to evolution of the signal along the line of sight. Although the coeval power spectrum is easy to compute in a simulation volume, it is not representative of a 3D power spectrum that would be observed. Given a flat $\Lambda$CDM cosmology, the comoving distance from an observer today can be calculated as a function of redshift:
\begin{equation}
r(z) = \int_0^z \frac{c}{H(z')} \dd{z'}.
\label{eqn:r(z)}
\end{equation}
As an example, if the center of the 2 Gpc/$h$ box is placed at a comoving distance corresponding to a redshift of $z=10$ for our particular cosmology (\textit{i.e.}, the 21 cm signal at the center of the box has a redshift of 10 relative to an observer), then a signal from the far side of the box (from the perspective of the observer) has a redshift of $z\sim 21$, whereas the near side of the box has a redshift of $z\sim 6$. The duration in redshift space spanned by the box is much larger than the $\Delta z_{50}$ for all of the reionization histories of our model. This means that, even for very extended reionization scenarios, the far side of the box would correspond to a totally neutral universe, and the near side would be completely ionized. The matter overdensity field also evolves from $z\sim 21$ to $z\sim 6$. Intuitively, one would expect that such a radical change could affect the power spectrum of 21 cm, because the signal is dependent upon the presence of neutral hydrogen. In other words, the evolution along the line of sight is non-negligible for these large volumes.

To produce the light cone effect, we divided the full simulation volume into a series of cubes with smaller dimensions, since 2 Gpc/$h$ spans a redshift range that always exceeds the duration of reionization in our model. We treated these different sub-boxes as fully independent, because the matter overdensity field, which generates the reionization field, has the same statistical values (\textit{e.g.}, mean value, standard deviation, $\sigma_8$, etc.) in each sub-volume, with some acceptable fluctuation. Specifically, we cut the 2 Gpc/$h$ box into sub-volumes of 500 Mpc/$h$, 250 Mpc/$h$, and 125 Mpc/$h$. This yields 64, 512, and 4096 independent cubes, respectively. We placed the center of the sub-boxes at the redshift corresponding to 25\%, 50\%, or 75\% neutral hydrogen fraction by mass. For each cell in the simulation volume, the comoving distance $r$ from the observer is calculated along with the redshift corresponding to that distance $z(r)$, the inverse of Equation~(\ref{eqn:r(z)}). Then, the mass of the cell is linearly interpolated from the snapshots of coeval mass density arrays from the bracketing redshifts, just as is done for the coeval case. Finally, the 21 cm brightness temperature is computed as in \S\ref{sec:ps}.

Figure~\ref{fig:lc-fields} shows the evolution of the 21 cm signal in the simulation volume as a function of redshift. One can see that the late-time portion (left side of the box) contributes almost nothing to the signal, and the earlier times (right side) has variation in the temperature proportional to the fluctuations in the matter overdensity field.

\subsection{3D Power Spectrum \\ with the Light Cone Effect}
\label{sec:3dlc}
To determine the impact the light cone effect has on the 3D power spectrum, we find the power spectrum of each individual sub-box, take the average, and then compute the standard deviation to get the corresponding 1$\sigma$ values. Because the simulation volumes were constructed in this way, periodicity was explicitly broken which altered the power on large scales. However, we found that this does not greatly affect the computation of the power spectrum. Furthermore, many of these results involve ratios between power spectra that are both affected by the problem of broken periodicity, so the problems introduced do not significantly change the predictions. Also note that when computing the 3D power spectrum with the light cone effect, Fourier modes where $k_\perp = 0$ relative to the line of sight have been removed. The inclusion of these modes leads to significantly more power on large scales, but they cannot be observed by radio interferometers. (See Appendix \ref{sec:exmodes} for more discussion.)

\begin{figure*}[p]
  \centering
  \includegraphics[width=0.95\textwidth]{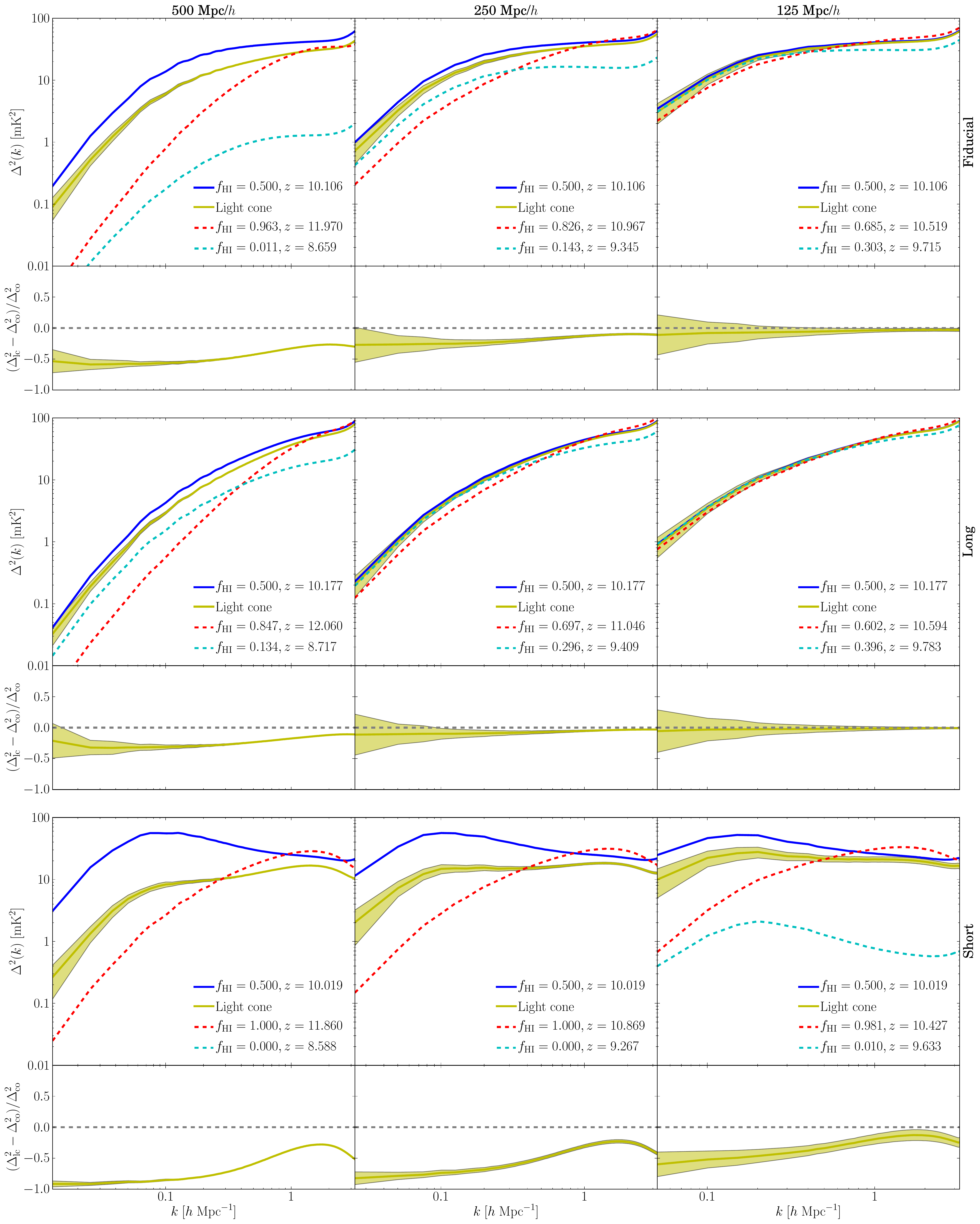}
  \caption{The light cone effect for a sub-box with side length 500 Mpc/$h$ (left column), 250 Mpc/$h$ (center column), and 125 Mpc/$h$ (right column), for the fiducial (top row), long (middle row), and short (bottom row) reionization scenarios. The coeval power spectrum (solid blue line) is computed at the midpoint of reionization. The light cone effect (yellow line) has 1$\sigma$ error regions shaded in. Note that these spectra do not include modes where $k_x = k_y = 0$ (see Appendix \ref{sec:exmodes}). Also shown are coeval power spectra corresponding to the bracketing redshifts of the light cone cube for the far side from the observer (red dashed line) and the near side (cyan dashed line). The percent difference between the coeval and light cone lines is shown in the bottom panel, with the same 1$\sigma$ error regions shaded in. The light cone effect is most pronounced at the largest scales. The light cone effect can also change the shape of the power spectrum, where a shorter reionization scenario leads to more deviation from the coeval case.}
  \label{fig:lc-ps-50}
\end{figure*}

In Figures~\ref{fig:lc-ps-50} and \ref{fig:lc-ps-fid} (along with Figures~\ref{fig:lc-ps-75}, \ref{fig:lc-ps-50-all} and \ref{fig:lc-ps-25} in Appendix~\ref{sec:more-figs}), we present the 3D power spectra with and without the light cone effect. Figure~\ref{fig:lc-ps-50} compares the power spectra across box sizes and reionization histories, but with constant $f_\mathrm{HI} = 0.5$. A general feature is that the power is suppressed at all scales. Including the light cone effect is somewhat analogous to averaging over the duration in redshift range spanned by the volume. For the large sub-volume size (500 Mpc/$h$), this leads to an effective averaging over a significant portion of the reionization history. This explains why there is less power on all scales: the neutral fractions where the large- and small-scale power peak ($f_{\mathrm{HI}}=0.50$ and $f_{\mathrm{HI}} = 0.75$, respectively) are being averaged with other neutral fractions that contain less power. Thus, the averaging tends to decrease power on all scales for our reionization scenarios. Note also that in the limit where the redshift space duration is relatively small (\textit{i.e.}, the 125 Mpc/$h$ volume), there is little deviation from the coeval case.

\begin{figure*}[p]
  \centering
  \includegraphics[width=0.95\textwidth]{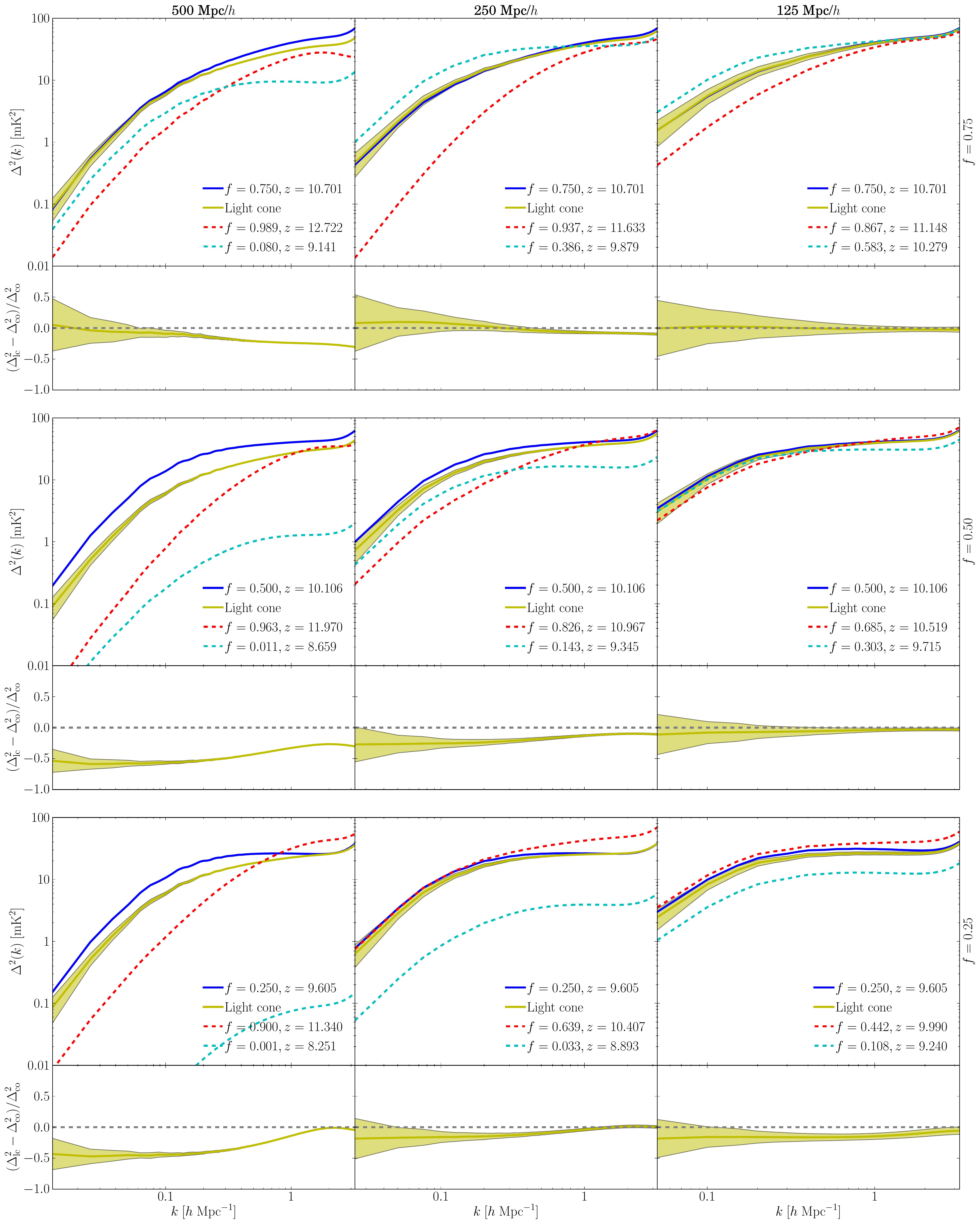}
  \caption{A plot similar to Figure~\ref{fig:lc-ps-50}, but showing the power spectrum as a function of neutral fraction. All plots are for the fiducial reionization history, with rows corresponding to $f_\mathrm{HI} = 0.75$ (top), 0.50 (center), and 0.25 (bottom). The columns have their same ordering as in Figure~\ref{fig:lc-ps-50}. We can see that only the small-scale power changes appreciably between different neutral fractions. Thus, on large scales, only the coeval power spectrum changes shape appreciably. Compare the coeval shape change to Figure~\ref{fig:3d-fid-ps}. This implies that the shape of the light cone power spectrum might not change as dramatically as in the coeval case, especially for large sample volumes.}
  \label{fig:lc-ps-fid}
\end{figure*}

\begin{figure*}[t]
  \centering
  \includegraphics[width=0.95\textwidth]{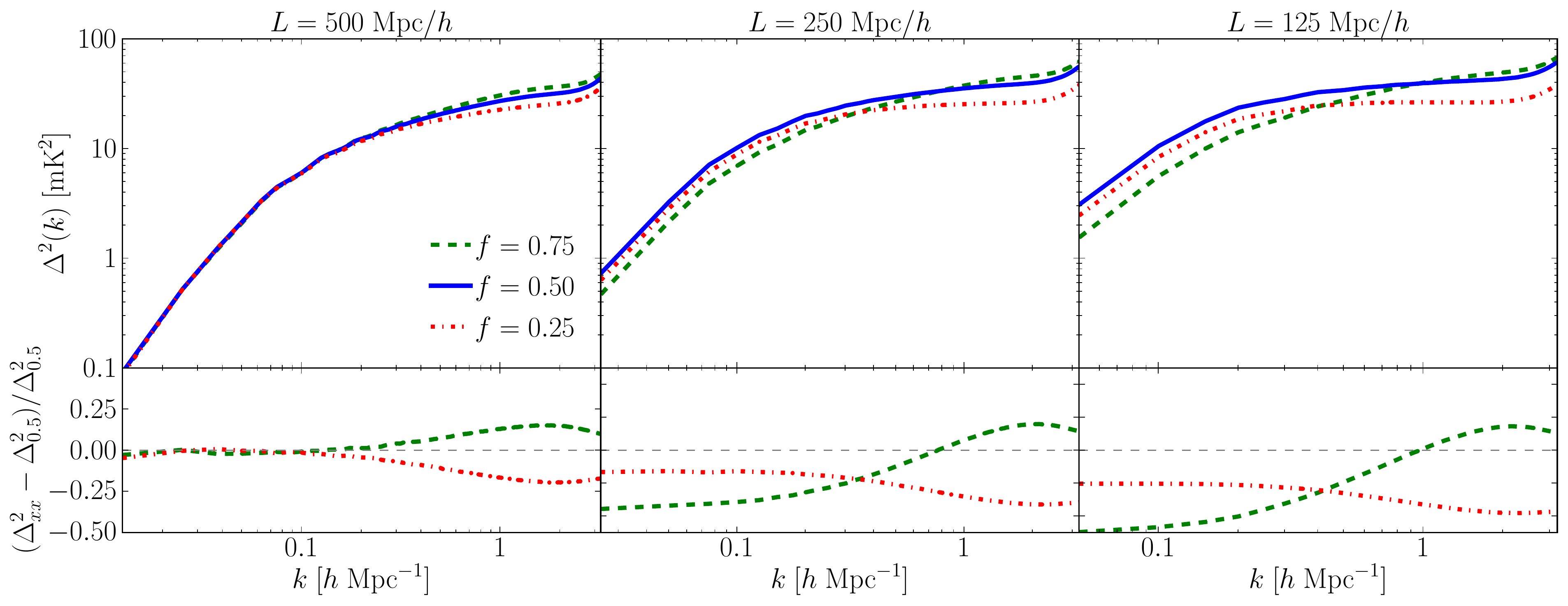}
  \caption{Top: light cone power spectra, plotted at different neutral fractions across all sub-box sizes for the fiducial reionization scenario. These spectra are the same as in Figure~\ref{fig:lc-ps-fid}, but reproduced here for more straight-forward comparison. Bottom: percent difference of the $f=0.75,0.25$ spectra from $f=0.50$ spectra. As the extent in redshift space becomes larger with respect to the duration of reionization, the large scale power becomes increasingly similar across neutral fractions. This effect is most apparent in the short scenario, but also partially seen in the long scenario.}
  \label{fig:lc-3d-fid}
\end{figure*}

One feature to point out is the 1$\sigma$ spread of the power spectrum, represented by a shaded region surrounding the light cone line. As discussed in \S\ref{sec:lc}, the sub-boxes are treated as independent and identically distributed sub-samples of the larger volume. Specifically, we treat the power spectrum from each sub-volume as the random variable of an underlying cosmological distribution. The standard deviation calculated here is that of the power spectra themselves, computed over the 64, 512, or 4096 sub-boxes for a particular sub-volume size. The relatively larger spread for the smaller sub-volumes demonstrates there is more fluctuation when examining smaller scales.

Figure~\ref{fig:lc-ps-fid} compares the 3D power spectrum across different box sizes and different neutral fractions, but only for the fiducial reionization scenario. Similar to Figure~\ref{fig:lc-ps-50}, the light cone enhances power at large scales and diminishes power at small scales. Another similarity is that the deviation from the coeval case is greater for large sub-volumes than for small ones. An interesting feature of these plots is that the shape of the light cone power spectrum does not change as drastically for different ionization fractions as it does for different reionization histories in our model. This implies that differences in the shape of the power spectrum are most sensitive to the duration of reionization, and are not as dependent on the midpoint of reionization.

Figure~\ref{fig:lc-3d-fid} shows the power spectra for the fiducial reionization scenario at different neutral fractions. As the sub-box size becomes larger and the extent in redshift space becomes large compared to the duration of reionization, the large-scale power of the different neutral fractions becomes increasingly similar. This is due to how the region of maximal contrast near $f_{\mathrm{HI}} \sim 0.5$ relates to where the box is centered in redshift space. Since the light cone cube is centered on the redshift corresponding to a particular neutral fractions, longer reionization scenarios will have a greater change in where the cubes are centered. Sub-box sizes where the region of maximal contrast is adequately spanned for all neutral hydrogen fractions will have similar amounts of large scale power. The Figure shows this is true for the largest sub-box size in the fiducial reionization scenario. As an observational implication, our model predicts that future measurements will not be able to easily distinguish different neutral fractions for briefer reionization scenarios.

Related to this phenomenon, the average bias (\textit{cf.}, \S\ref{sec:cc}) also behaves differently when the light cone effect is included. In the coeval case, the average bias is initially relatively small early in reionization, rises with increased ionization, and then falls following the midpoint of reionization (see Figure~\ref{fig:bias}). The inclusion of the light cone effect flattens out this curve, so that the average bias does not change significantly as a function of neutral fraction. Again, this phenomenon is related to the duration of reionization compared to sub-box size, with briefer reionization scenarios being flatter. This result further demonstrates that when the light cone effect is included, it becomes difficult to determine the change in neutral fraction as a function of redshift. One alternative to the 3D power spectrum would be to measure the 2D angular power spectrum as a function of frequency, where the large scale bias would likely rise and fall as a function of neutral fraction in a manner similar to the coeval case.

\subsection{Cross-correlation Coefficient \\ with the Light Cone Effect}
\label{sec:cclc}
\begin{figure*}[t]
  \centering
  \resizebox{0.5\hsize}{!}{\includegraphics{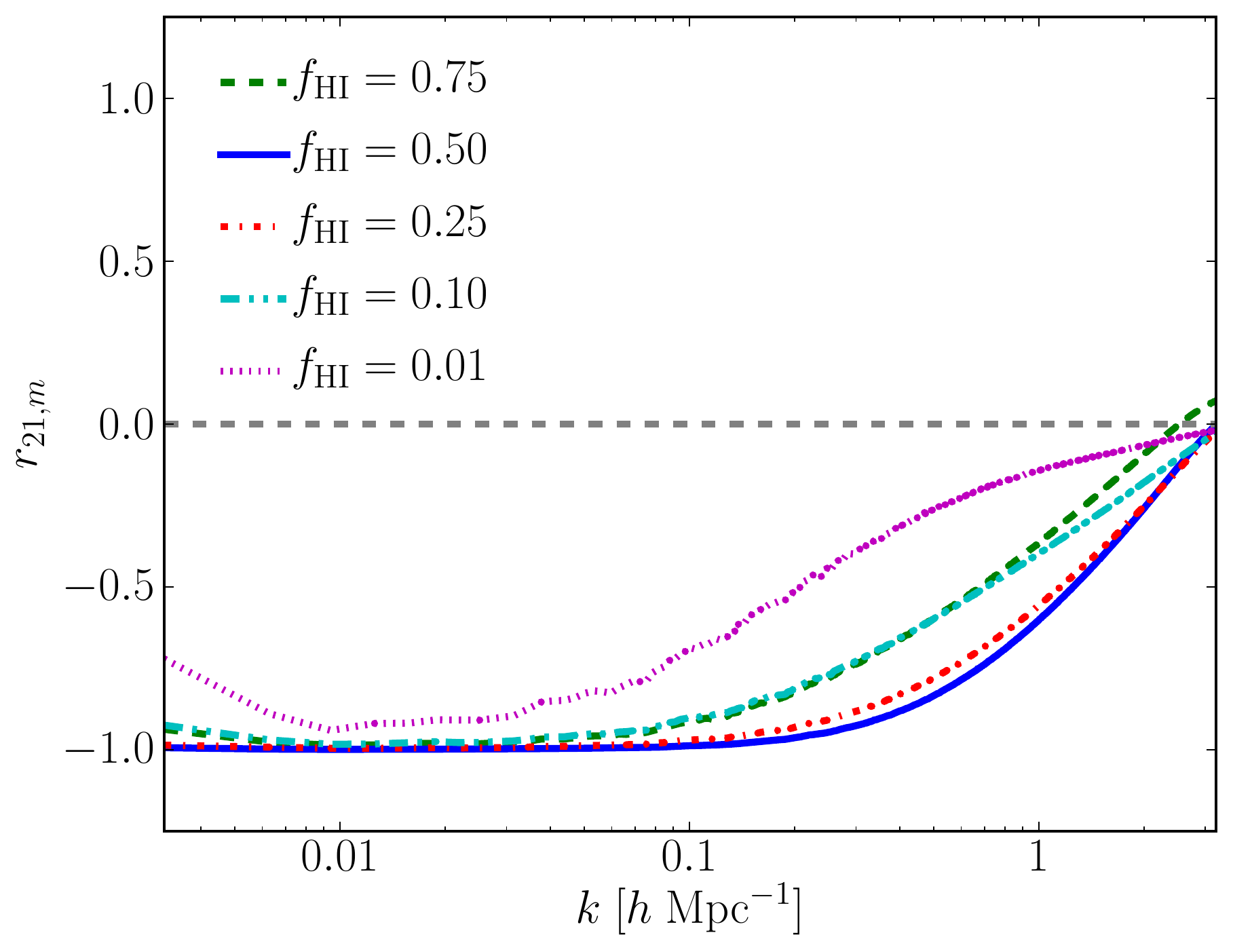}}%
  \resizebox{0.5\hsize}{!}{\includegraphics{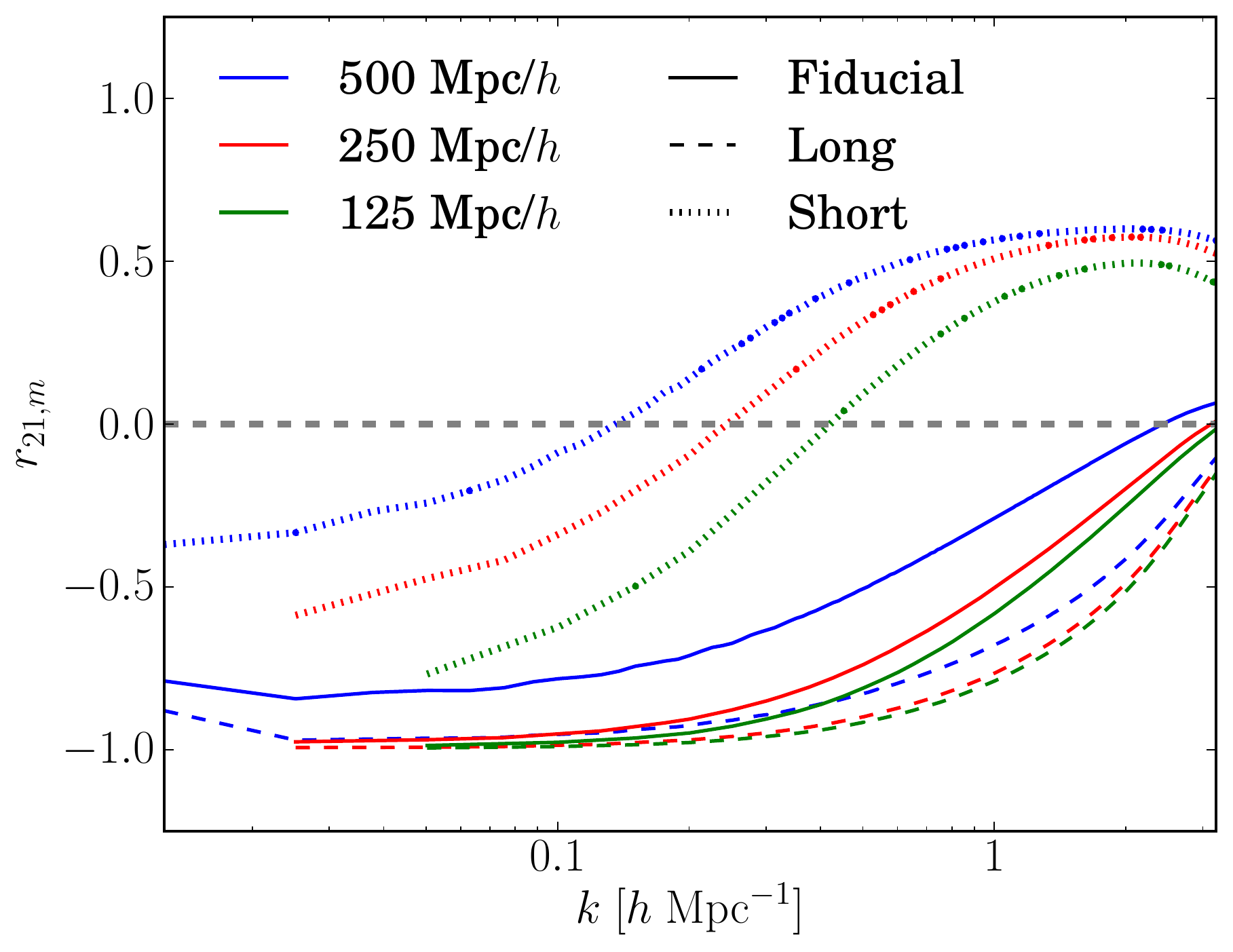}} \\
  \caption{Left: the cross-correlation coefficient between the 21 cm brightness temperature and the matter overdensity field, defined in Equation~(\ref{eqn:cc}), for the fiducial reionization history plotted at different neutral fractions. On very large scales, there is almost perfect anti-correlation between the two fields. Right: cross-correlation coefficient including the light cone effect, across different sub-box sized and reionization scenarios at constant $f_\mathrm{HI} = 0.5$. For the fiducial and long reionization scenarios, there is generally a tendency toward $-1$ on large scales, though the anti-correlation is not as pronounced as in the coeval case. However, this correlation does not exist to the same extent for the short reionization scenario.}
  \label{fig:corr}
\end{figure*}

We examined the cross-correlation coefficient between the 21 cm brightness temperature and the matter overdensity fields for the light cone effect. We computed the cross-correlation between the two fields using Equation~(\ref{eqn:cc}). Figure~\ref{fig:corr} shows the cross-correlation coefficient for the light cone. In general, on large scales the fields show less statistical correlation than in the coeval case. We can motivate this by noting that when the box is completely neutral, there is perfect correlation between the two fields. Conversely, once the box becomes totally ionized, there is no longer any correlation between the matter overdensity and 21 cm fields, because the 21 cm signal is zero everywhere. Because this effect is more pronounced in the short reionization scenario (\textit{cf.} Figure~\ref{fig:lc-fields}), the short histories (the dotted lines in Figure~\ref{fig:corr}) deviate the most from perfect anti-correlation. In fact, the combination of zero correlation in ionized regions and almost perfect correlation in neutral regions accounts for why the short reionization scenario exhibits a large degree of positive correlation on small scales. The amount of anti-correlation grows larger for longer reionization scenarios, and the fields tend toward perfect anti-correlation on large scales for the fiducial and long reionization scenarios.

\subsection{Anisotropic Power Spectrum}
\label{sec:2dps}

We are interested in quantifying any anisotropy in the power spectrum because the light cone effect inherently alters the signal along the observer's line of sight of the volume, but does not affect the signal perpendicular to the line of sight. The computation of an anisotropic power spectrum proceeds in a fashion similar to that of the 3D power spectrum (as in \S\ref{sec:ps}); however, instead of binning in terms of a single spherical magnitude $k = \sqrt{\smash{k_x}^2 + \smash{k_y}^2 + \smash{k_z}^2}$, the binning is done in terms of two quantities $k_\parallel \equiv k_z $ and $k_\perp \equiv \sqrt{\smash{k_x}^2 + \smash{k_y}^2}$. When decomposing the power spectrum in this manner, we use a ``flat-sky'' approximation which neglects the curvature of the sky. In our calculations, the distance to the observer is large enough that the effects of the flat-sky approximation are negligible. Additionally, we noticed that on small scales, there is a significant amount of anisotropy in the Figure even in the coeval matter power spectrum. The density field is constructed by assigning particles to a Cartesian grid using an anisotropic cubical top hat filter. Deconvolution with this filter is not perfect, and does not completely remove the anisotropy. The deviation from isotropy becomes increasingly important on scales that are close to the size of a grid cell. Accordingly, we only trust this statistic for which $k\lesssim 1$ $h$ Mpc$^{-1}$.

\begin{figure*}[t]
  \begin{minipage}[t]{0.47\hsize}
    \centering{\small Coeval:}
  \end{minipage}
  \begin{minipage}[t]{0.47\hsize}
    \centering{\small Fiducial:}
  \end{minipage}
  \centering{
    \resizebox{0.47\hsize}{!}{\includegraphics{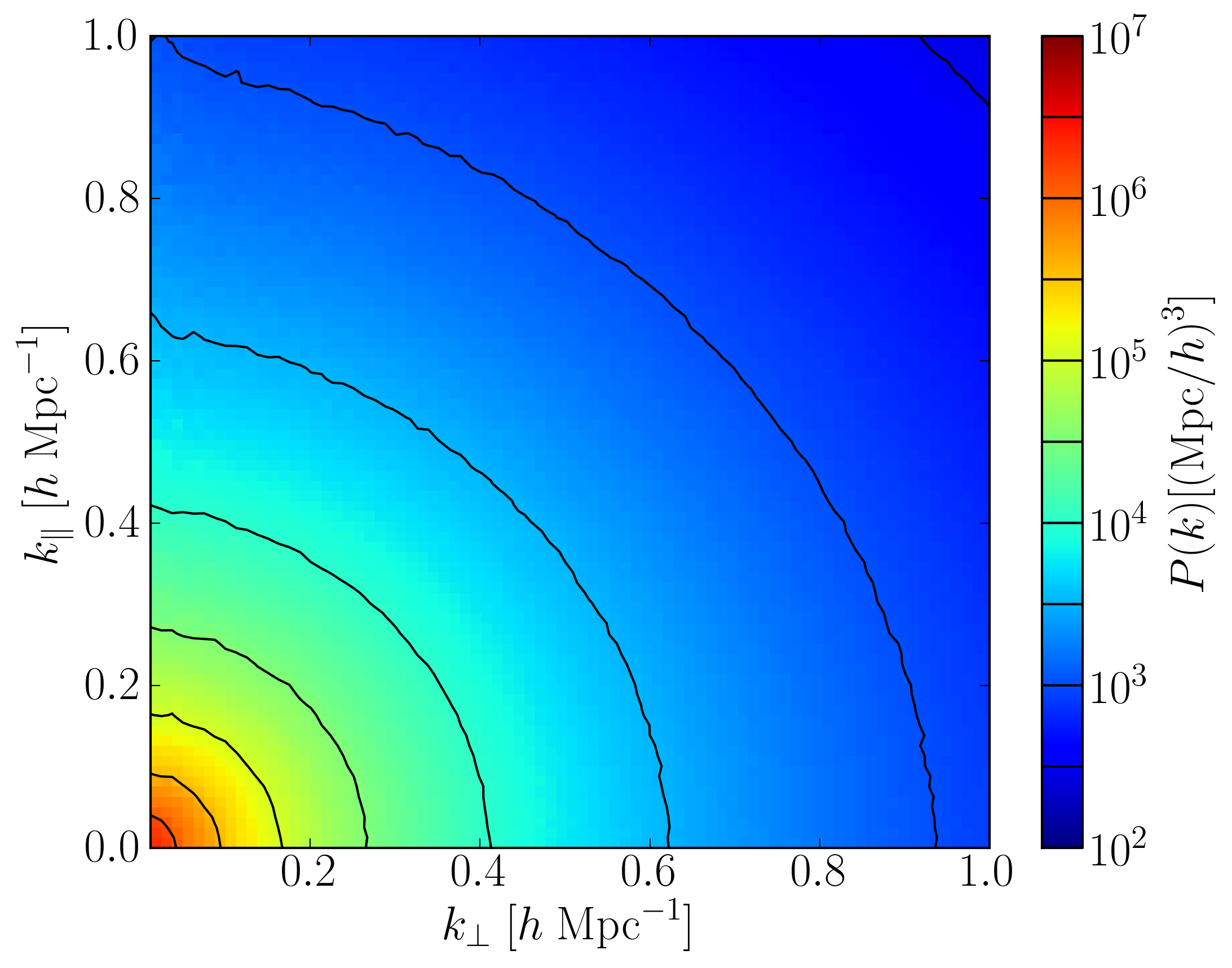}}%
    \hspace{2pt}%
    \resizebox{0.47\hsize}{!}{\includegraphics{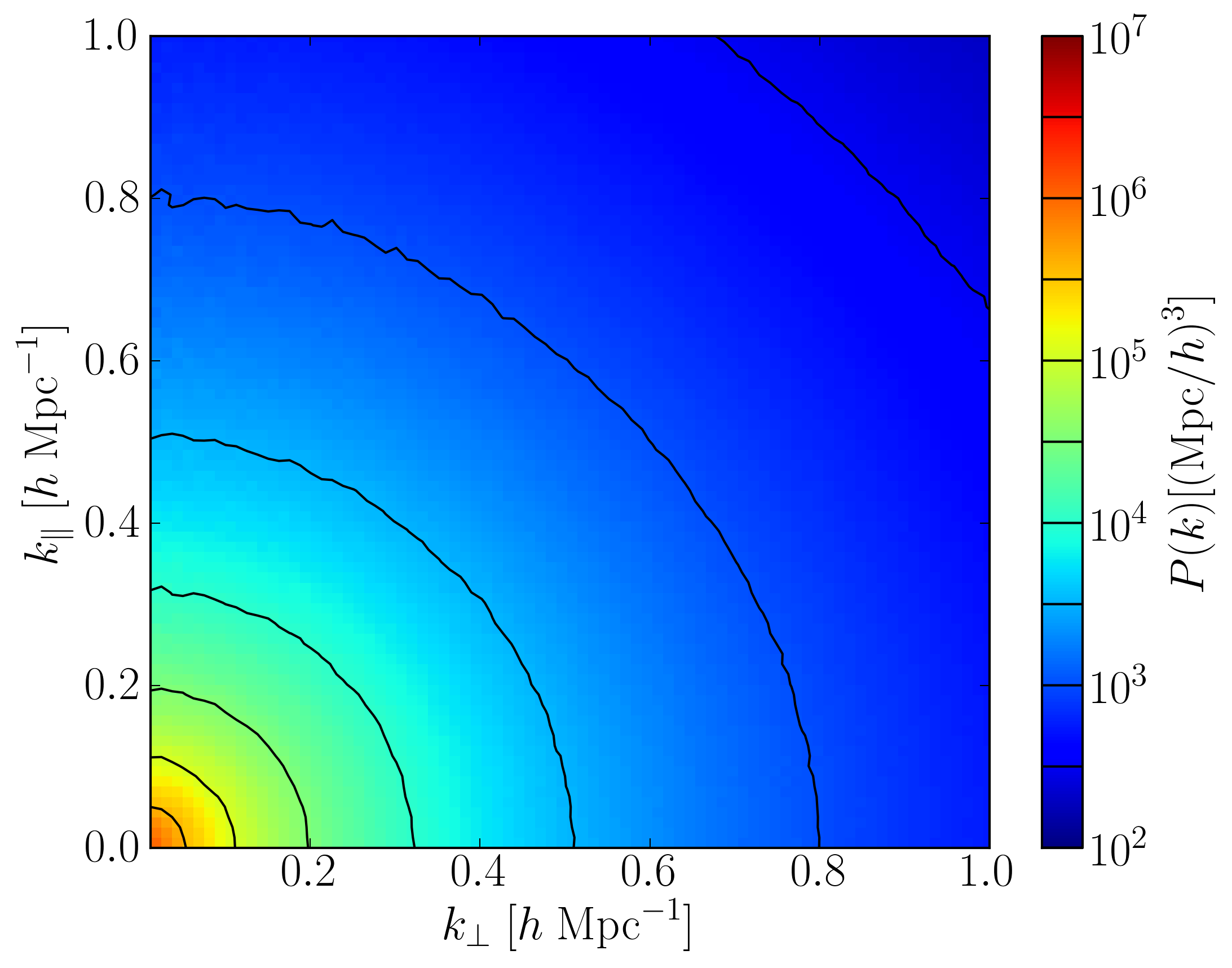}} }\\
  \begin{minipage}[t]{0.47\hsize}
    \centering{\small Long:}
  \end{minipage}
  \begin{minipage}[t]{0.47\hsize}
    \centering{\small Short:}
  \end{minipage}
  \centering{
    \resizebox{0.47\hsize}{!}{\includegraphics{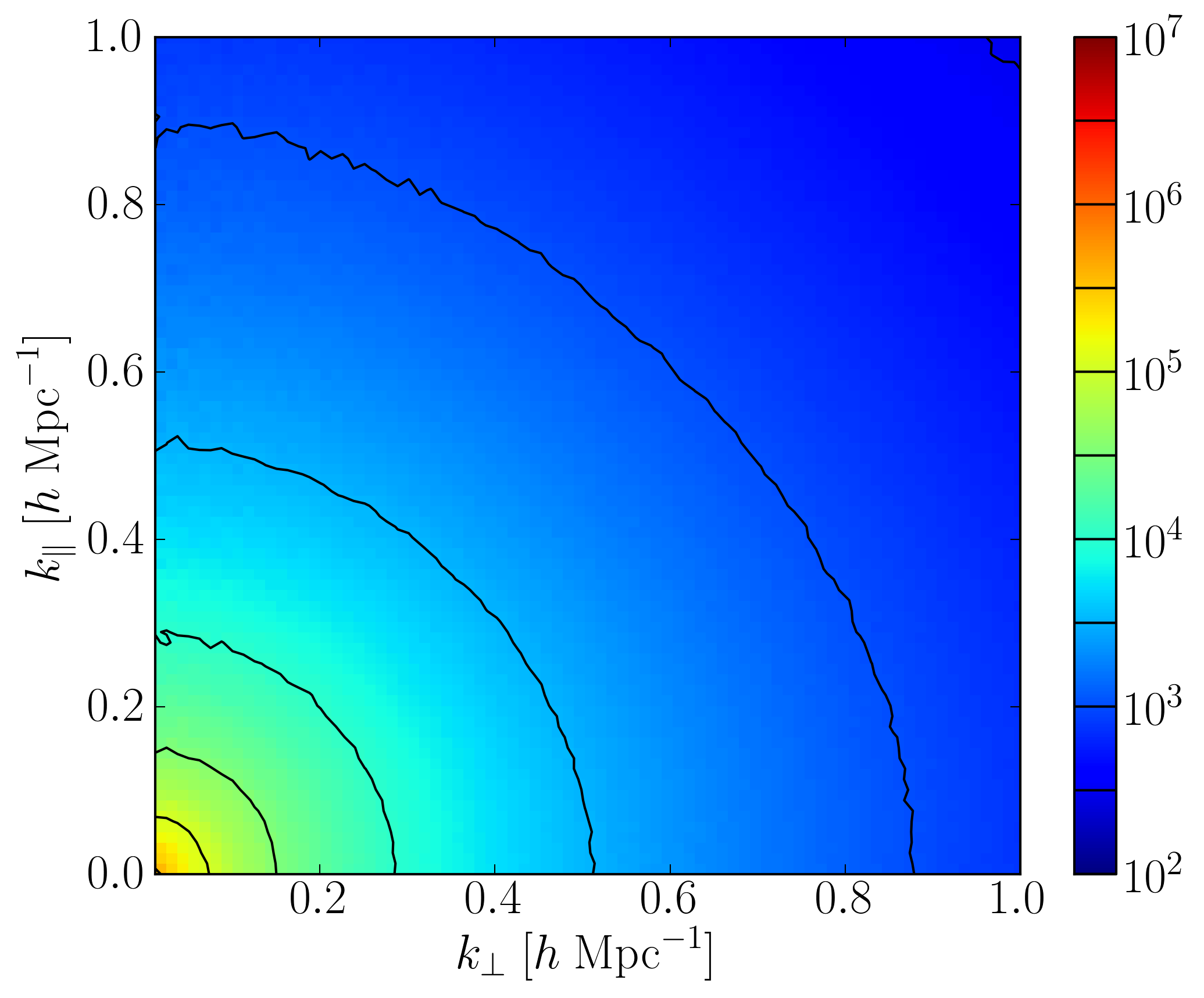}}%
    \hspace{2pt}%
    \resizebox{0.47\hsize}{!}{\includegraphics{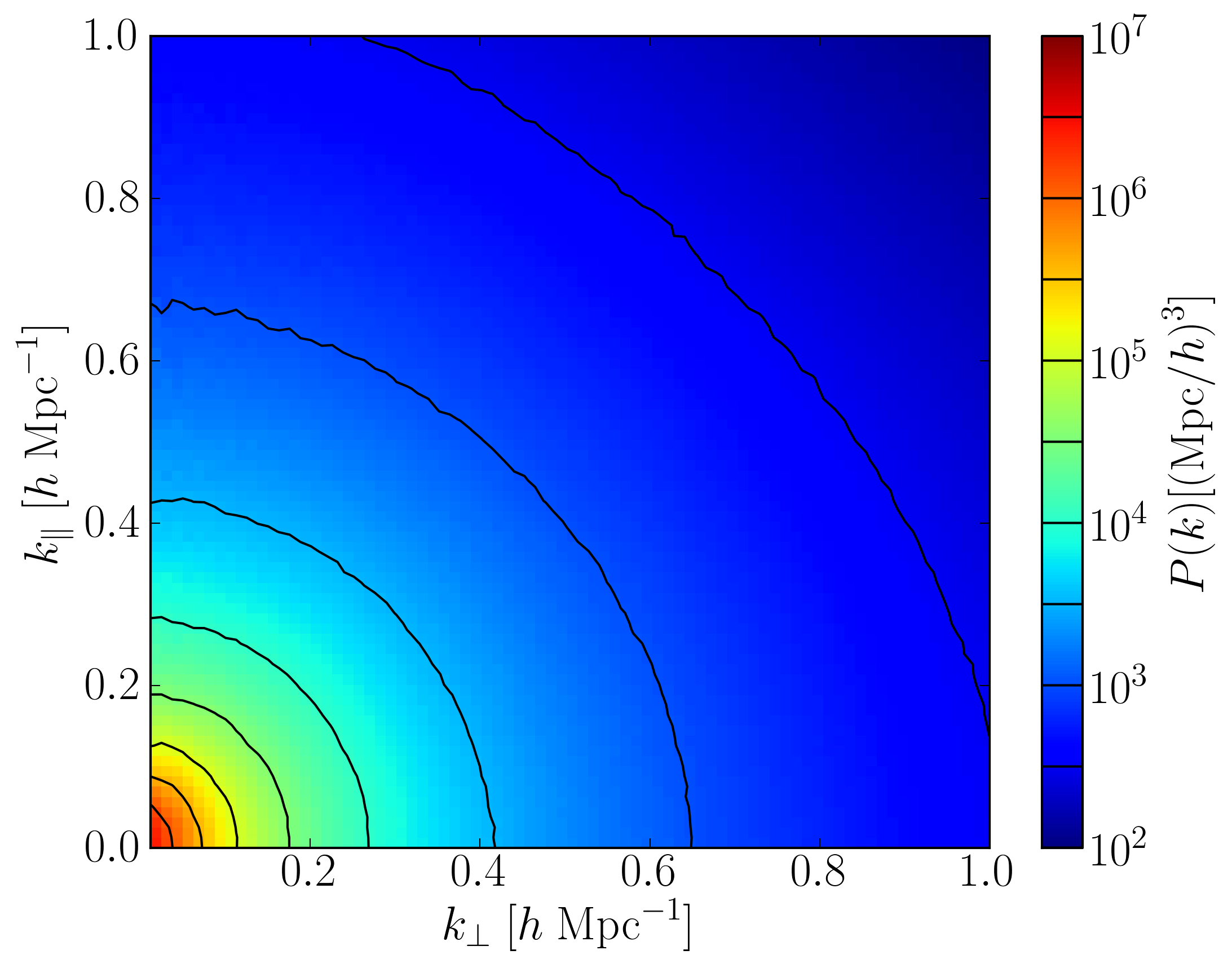}} }\\
  \caption{A plot of the anisotropic power spectrum, broken down into parallel and perpendicular Fourier modes. Top left: the anisotropic power spectrum of the coeval 21 cm field, $f_\mathrm{HI}=0.5$, fiducial reionization scenario. As with the 3D power spectrum, the result has been averaged over the 64 independent 500 Mpc/$h$ sub-boxes. Top right: same plot, but including the light cone effect and all Fourier modes for the fiducial reionization scenario. Here, $k_{\parallel}$ is taken to be along the line of sight and coincident with the direction of the light cone effect. Both cases appear similarly isotropic. The anisotropy changes slightly based on the reionization history, especially in the short case. The effect is also more pronounced for larger scales. Bottom: the long and short reionization scenarios, respectively.}
  \label{fig:2dps-fid}
\end{figure*}

Figure~\ref{fig:2dps-fid} shows a pseudo-color plot in which the different $k$-modes $k_\perp$ and $k_\parallel$ are on the $x$- and $y$-axes, respectively. The power spectrum $P(k)$ is plotted as a function of these two modes on a linear scale, so that the isotropy (or anisotropy) is apparent in the plot. An interesting feature to point out is that the light cone introduces a subtle deviation from the isotropy seen in the coeval case. There slightly less power in modes where $k_\perp \sim k_\parallel$ compared to modes where $k_\perp \gg k_\parallel$ or vice versa. Compare this to the coeval case, where the contours are almost perfectly circular with little deviation from isotropy. The anisotropy indicates there is more power for volumes with small extent in redshift space or small extent in the plane of the sky, compared to ones where the extent is almost equal.

In the case of an isotropic box with no preferred direction (\textit{e.g.}, a coeval cube containing the matter overdensity field), one would expect the contours of equal power to be roughly circular, because there should be equal contributions in all directions without a preferred orientation. When the light cone effect is included, we find that there is generally less power at all scales $k \lesssim 1$ $h$ Mpc$^{-1}$, which is consistent with Figure~\ref{fig:lc-ps-50}. Figure~\ref{fig:lc-ps-50} demonstrates that including the light cone effect leads to a similar spectrum but with less power at all scales, and Figure~\ref{fig:2dps-fid} shows that there is little anisotropy introduced by the effect.

\subsection{Power Wedges}
\label{sec:wedges}

We quantify the anisotropy produced by the light cone effect using a tool we name ``power wedges,'' in analogy to the ``clustering wedges'' tool recently introduced in BAO analysis for the two-point correlation function \citep[\textit{e.g.},][]{kazin_etal2013,sanchez_etal2013}. To perform the power wedges analysis, the plane of $k_\parallel$ and $k_\perp$ is bisected along the line $k_\parallel = k_\perp$. Then, the power corresponding to these combinations is binned as a function of $k$. This process produces decompositions $P_\parallel$ and $P_\perp$, where $k_\parallel > k_\perp$ or vice versa. Finally, the ratio of the power spectra ($\chi(k)$) is taken:
\begin{equation}
\chi(k) \equiv \frac{\ev{P_\parallel (k)}}{\ev{P_\perp (k)}}.
\label{eqn:chi}
\end{equation}
In the case that the $k$-values are equal, the contribution to the power is added to both spectra. In a perfectly isotropic case, this parameter should be equal to 1 (with some fluctuation). If the parameter is greater than 1, then there is more power coming from the modes along the line of sight of the simulation box, and vice versa.

\begin{figure*}[t]
  \centering
  \includegraphics[width=\textwidth]{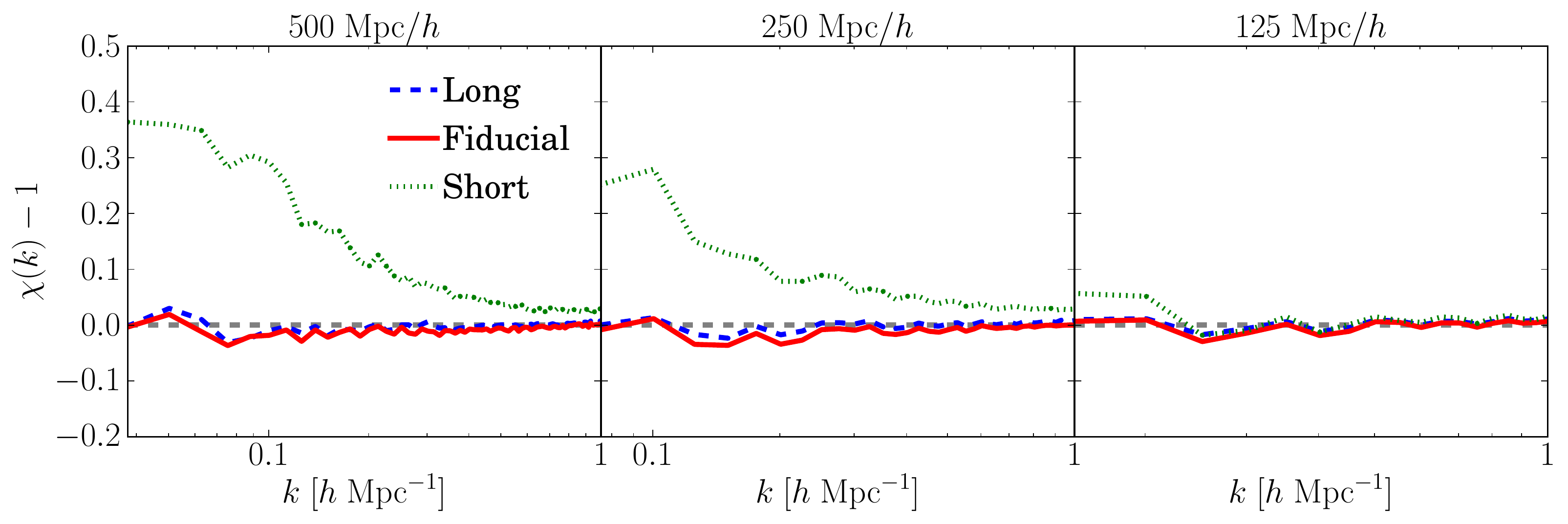}
  \caption{The power wedges measurement, across different sub-box size and reionization histories. The fiducial and long reionization scenarios show little anisotropy.  However, for larger sub-box sizes and on large scales, the short reionization scenario does display some anisotropy. This is due to the fact that for the short reionization scenarios, there is greater variance in the 21 cm signal along the line of sight compared to the longer reionization scenarios. Note that as the scales considered become smaller, all of the scenarios are nearly isotropic.}
  \label{fig:wedges-fid}
\end{figure*}

Figure~\ref{fig:wedges-fid} shows the results of using the power wedges analysis. One can see that $\chi$ changes noticeably as a function of reionization history. For the fiducial and long histories, the value is very close to 1, meaning that the signal is isotropic. However, the short history demonstrates a moderate degree of anisotropy on large scales (almost $\sim40\%$). The deviation from 1 becomes less as the sub-box size becomes smaller. Physically, the shorter reionization scenario displays a greater change in the variance of the 21 cm signal along the line of sight compared to the long reionization scenario.

Another interesting result evident in Figure~\ref{fig:wedges-fid} is how the evolution within the volume affects $\chi(k)$. There is a much larger deviation from unity for the case of the short reionization scenario compared to the fiducial and long ones. For a larger sub-box size, there is more evolution of the neutral hydrogen fraction, especially for the short reionization scenario. Because the anisotropy induced depends on this evolution, the larger anisotropy for larger box sizes makes sense.

\subsection{Comparison to Previous Work}
\label{sec:prev}
As mentioned previously, the light cone effect has been investigated in \citet{datta_etal2012}. The work presented here differs from the previous one in several key aspects. First, some of the volumes considered here are significantly larger. The simulation volume in the previous work was 163 Mpc ($\approx$114 Mpc/$h$) on a side, compared to the light cone sub-box volumes of 500, 250, and 125 Mpc/$h$. In the previous work, the light cone was predicted to deviate from the coeval signal by $\sim$30-40\%. Additionally, the previous work also found for the early- and mid-points of reionization, there was an increase in power when compared to the coeval case on large scales, and a decrease at small scales. When looking at the results for the 125 Mpc/$h$ sub-box, we find that the predictions presented here match the ones presented previously, but only in the long reionization scenario early in reionization. Since our long reionization duration is comparable to their fiducial case, there is good agreement. We also note that the light cone effect becomes increasingly important as the scales get larger. As Figure~\ref{fig:lc-ps-50} shows, for the 500 Mpc/$h$ volumes, the light cone effect can deviate by more than 50\% for the fiducial scenario and up to an order of magnitude for the short scenario. To see the full effect of the light cone, larger volumes must be used.

Another difference is that the light cone cubes presented here are constructed from a sub-volume of the entire simulation volume available. In the previous work, the light cone volume was the same size as the total simulation volume. This leads to pseudo-periodic boundary conditions in the perpendicular directions. Breaking the periodicity of the FFT can have important implications on the predicted power spectrum, especially for large-scale modes. These considerations are especially important for real-world data acquisition, where in general periodic boundary conditions do not apply. So, by explicitly breaking periodicity with the light cone cubes, we present predictions that will more readily conform to practical data processing.

The use of sub-volumes in the light cone calculation also means we are able to eliminate much of the cosmic variance for large scales. By averaging the power spectra over many independent sub-volumes of the total simulation volume, we reduce the scatter inherent in the large scale modes. Accordingly, we are able to make progress toward a smooth power spectrum, creating an improved statistical measure of the 21 cm brightness temperature field.

\section{Observational Comparison}
\label{sec:observation}
Recently, upper limits on the 21 cm signal were derived based on data from the Precision Array for Probing the Epoch of Reionization (PAPER) \citep{parsons_etal2010,pober_etal2013}. Specifically, we are interested in recent results presented in \citet{parsons_etal2013}, which reported an observational upper-limit on the 21 cm power spectrum of 2700 $\mathrm{(mK)}^2$ at a redshift of $z = 7.7$ in the neighborhood of $k \sim 0.1$ $h$ Mpc$^{-1}$. We computed a predicted observation using the bias model discussed in \S\ref{sec:methodology}. We considered here a prediction for the light cone power spectrum, with a midpoint of reionization to be $\bar{z}=8$ for a more apt comparison, using the 500 Mpc/$h$ sub-box size, measured at $f_\mathrm{HI} = 0.5$ by volume (which corresponds to $z = 7.9$).

The Giant Metrewave Radio Telescope (GMRT) also has derived upper limits on the 21 cm signal from measurements \citep{paciga_etal2011,paciga_etal2013}. In this result, GMRT has upper limits on the power spectrum amplitude at a redshift of $z=8.6$ in the neighborhood of $k \approx 0.50$ $h$ Mpc$^{-1}$. The most restrictive measurement at 2$\sigma$ is (248 mK)$^2$ at $k$ = 0.50 $h$ Mpc$^{-1}$, with 4 singular value decomposition (SVD) modes removed to correct for foreground contamination. (See \citealt{paciga_etal2013} for further explanation.) One aspect to note is that the foreground removal techniques of PAPER and GMRT are different, and the measurements are reported for different redshift values. A direct comparison should not be made between the two, but instead compared directly to the theoretical prediction (solid line).

\begin{figure}[t]
  \centering
  \includegraphics[width=0.45\textwidth]{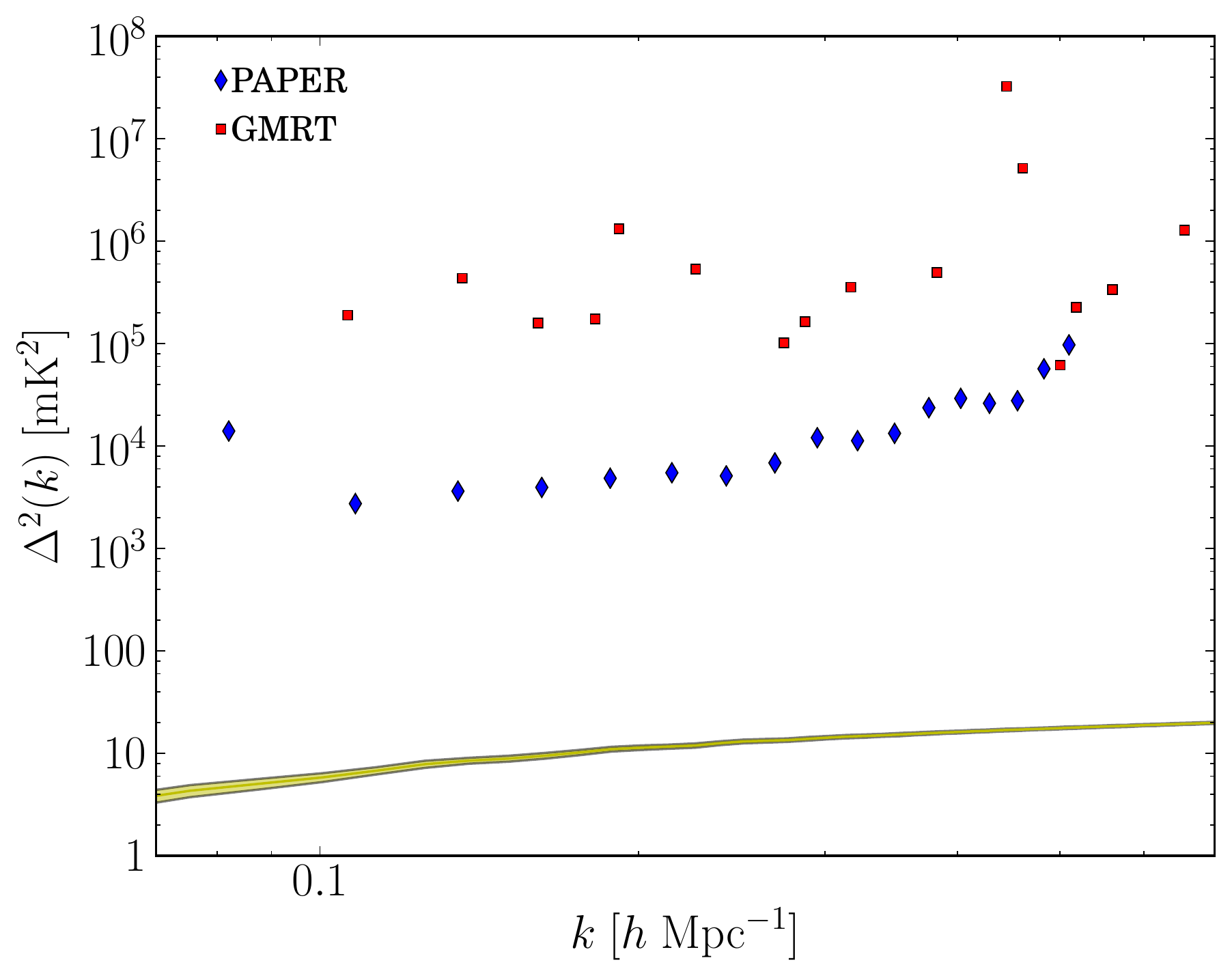}
  \caption{A comparison of experimental results from PAPER \citep{parsons_etal2013} and GMRT \citep{paciga_etal2013} with theoretical predictions incorporating the light cone. The data from PAPER represents 2$\sigma$ upper limits. The data from GMRT also represents 2$\sigma$ upper limits of the power spectrum. The solid curves are the predicted power spectrum of the 500 Mpc/$h$ sub-box for the fiducial reionization scenario at 50\% ionization with a midpoint of reionization at $\bar{z}=8$. The difference between the predictions and the data is several orders of magnitude.}
  \label{fig:real-data}
\end{figure}

Figure~\ref{fig:real-data} presents the 21 cm power spectrum upper-limits from PAPER \citep{parsons_etal2013} and GMRT \citep{paciga_etal2013}, compared to the reionization model at 50\% reionization for $\bar{z}=8$ with the light cone effect. For the plot of GMRT data, we selected the most restrictive point among the different number of SVD modes removed. The predicted amplitude is $\sim10-100$ mK$^2$, which is at least two orders of magnitude smaller than the upper limits reported by PAPER and GMRT. However, other theoretical predictions that do not include exotic reionization scenarios have similar order of magnitude differences \citep[\textit{e.g.},][]{zahn_etal2007,iliev_etal2008}. Varying the reionization history did not raise the signal to the same order of magnitude of the upper limits. 

Another important observational constraint comes from the Experiment to Detect the Global EoR Step (EDGES) experiment \citep{bowman_rogers2010}. In this result, the authors reported a lower limit to the duration of reionization, stating that the total duration of reionization is $\Delta z_{50} \gtrsim 0.07$ with 95\% confidence. We have converted the EDGES definition of $\Delta z$, which assumes a functional form of a hyperbolic tangent, to the definition of $\Delta z_{50}$ discussed in \S\ref{sec:21cm}. The short reionization scenario has a 50\% reionization duration of $\Delta z_{50} = 0.24$. Thus, the EDGES observations do not yet rule out any of the theoretical models presented here.

\section{Discussion}
\label{sec:discussion}
An important observational consideration when measuring the 21 cm signal is the process of foreground removal. 21 cm brightness temperature fluctuations are typically 3-5 orders of magnitude smaller than signals coming from foreground contamination, such as galactic synchrotron radiation and extragalactic point sources. Typical schemes for removing these contaminants are to look at their spectra in frequency space. The 21 cm signal is expected to vary rapidly as a function of frequency, whereas these contaminants are expected to vary smoothly \citep{zaldarriaga_etal2004,mcquinn_etal2006b,liu_etal2009}. By removing these smoothly varying components from the spectrum, the true 21 cm signal emerges from the foregrounds. Unfortunately, this technique may also remove some of the long-frequency modes of the power spectrum, which is also the region of interest for the light cone effect. Care must be taken to ensure that 21 cm signal is not being discarded along with the foregrounds.

The implications of the light cone effect can also be compared to the effect of redshift space distortions (RSD). Recent work by \citet{jensen_etal2013} showed that RSD are most important at early stages of reionization ($0.7 \lesssim f_\mathrm{HI} \lesssim 1.0$). At these stages, RSD contribute to an $\order{1}$ enhancement of the 3D power spectrum on large scales $k \lesssim 0.3\ h$ Mpc$^{-1}$. For later stages in reionization, RSD have a less important effect, and by $f_\mathrm{HI} \sim 0.5$ onwards, RSD induce only a percent-level change on the 3D power spectrum. As Figure~\ref{fig:lc-ps-fid} demonstrates, the light cone can have a decrement of up to $\sim 50\%$ for scales $k \lesssim 0.02\ h$ Mpc$^{-1}$. Figure~\ref{fig:lc-ps-50} shows that the light cone effect has an important effect on large scales for the midpoint of reionization, $0.25 \lesssim f_\mathrm{HI} \lesssim 0.75$. RSD also introduce an anisotropy to the 3D power spectrum, though the formalism presented in \citet{jensen_etal2013} expresses the anisotropy as an expansion in terms of $\mu \equiv \cos \theta$, the angle between the line of sight and the direction in $k$-space. We plan to further investigate the different implications of the light cone effect and RSD in future work.

\section{Conclusions}
\label{sec:conclusion}
We accomplished the following in this paper:
\begin{itemize}
\item Using a parametrized bias factor between the redshift of reionization and the matter overdensity field, we created a reionization field for a large ($\sim$2~Gpc/$h$) simulation volume.
\item We made predictions about the global 21 cm brightness signal using this large volume.
\item We calculated the 3D power spectrum and cross-correlation coefficient for both the coeval and light cone cases.
\item We showed that including the light cone effect makes a moderate difference in the amplitude (up to 50\% for small $k$-modes), and can change the shape of the spectrum at all scales.
\item Using ``power wedges'' analysis, we showed that the anisotropy introduced by the light cone is only present for our short reionization scenario. We also showed this anisotropy is most sensitive to large changes in the neutral fraction of the contained volume. Thus, the light cone effect likely will not induce significant anisotropy in upcoming experiments.
\item We compared predictions from our model to the recent results from the PAPER and GMRT surveys, and showed that our predictions are an order of magnitude smaller than their upper-limits on the 3D power spectrum of the 21 cm brightness temperature signal.
\end{itemize}

As mentioned in \S\ref{sec:intro}, the light cone effect has important implications for measurements that use the BAO method. The BAO scale, $\sim$150 comoving Mpc ($k \sim 0.06$ $h$ Mpc$^{-1}$), approaches the scale where the light cone effect becomes non-negligible. The light cone effect can have up to a $\sim50\%$ effect on the predicted signal on these scales. We have also shown that the light cone effect can introduce an anisotropy along the line of sight for short reionization scenarios. This complicates using the Alcock-Paczy\'{n}ski test to determine the proper cosmological parameters of the universe. Future applications of the BAO method to the 21 cm signal will have to account for the light cone effect in their analyses.

In future work, we would like to include redshift space distortions with the light cone effect. RSD have been investigated with respect to the 21 cm signal \citep[\textit{e.g.},][]{bharadwaj_ali2004,barkana_loeb2005,bharadwaj_ali2005,mao_etal2012,jensen_etal2013,shapiro_etal2013,majumdar_etal2013}. However, these previous explorations did not include the light cone effect in their analysis. We would like to examine both simultaneously, and determine which scales are important for the effects, and how measurements are affected by each. As the data thus far suggests, smaller volumes are affected less by the light cone effect; the logical conclusion of this observation would be to analyze the 2D power spectrum, where we only examine the signal in the plane of the sky for a very narrow redshift range. By performing the analysis in this fashion, we are no longer plagued by the problem of disproportionate power from along the line of sight, but we potentially lose out on valuable three dimensional information. Thus, we hope to make predictions at different points in redshift/frequency space and then combine the results to reconstruct the 3D signal.

\acknowledgments{N. B. and A. N. are supported by a McWilliams Center for Cosmology Postdoctoral Fellowship made possible by Bruce and Astrid McWilliams. A.N. and J.B.P. acknowledge funding from NSF grant AST-1009615. We thank A. Parsons for supplying the data from PAPER and G. Paciga for supplying the data from GMRT. H. T. is supported in part by NSF grants AST-1109730 and AST-1312991. R. C. is supported in part by NSF grant AST-1108700 and NASA grant NNX12AF91G. A. L. is supported in part by NSF grant AST-0907890 and NASA grants NNX08AL43G and NNA09DB30A. The simulations were performed at the Pittsburgh Supercomputing Center (PSC) and the Princeton Institute for Computational Science and Engineering (PICSciE). We thank Roberto Gomez and Rick Costa at the PSC and Bill Wichser at PICSciE for invaluable help with computing.}

\appendix

\section{Exclusion of $ \lowercase{k}_\perp = 0$ Modes}
\label{sec:exmodes}
Modes where $k_\perp = 0$ correspond to the total flux at a particular frequency defined by $k_\parallel$. For radio interferometers, this mode is inaccessible, since interferometers only measure fluctuations relative to a background level. Alternatively, to probe modes where $k_\perp = 0$, the antennas would have to have no separation between them, which is not possible. These modes would not be detectable in most experiments proposing to measure the 21 cm brightness temperature \citep{datta_etal2012}.

In order to determine how the exclusion of the $k_\perp = 0$ modes changed our predictions, we performed the preceding analysis both including and excluding these modes. Removing these modes is roughly equivalent to subtracting the mean temperature from each 2D slice in the $xy$-plane. Accordingly, the variance measured by the power spectrum has three components: the change in the average neutral fraction, the change in this average temperature as a function of redshift, and the average HII region bubble size as a function of redshift. The removal of $k_\perp = 0$ essentially eliminates the variance due to the changing average temperature, but it does not eliminate the contributions from changing neutral fraction contribution or the bubble size.

\begin{figure*}[p]
  \centering
  \includegraphics[width=\textwidth]{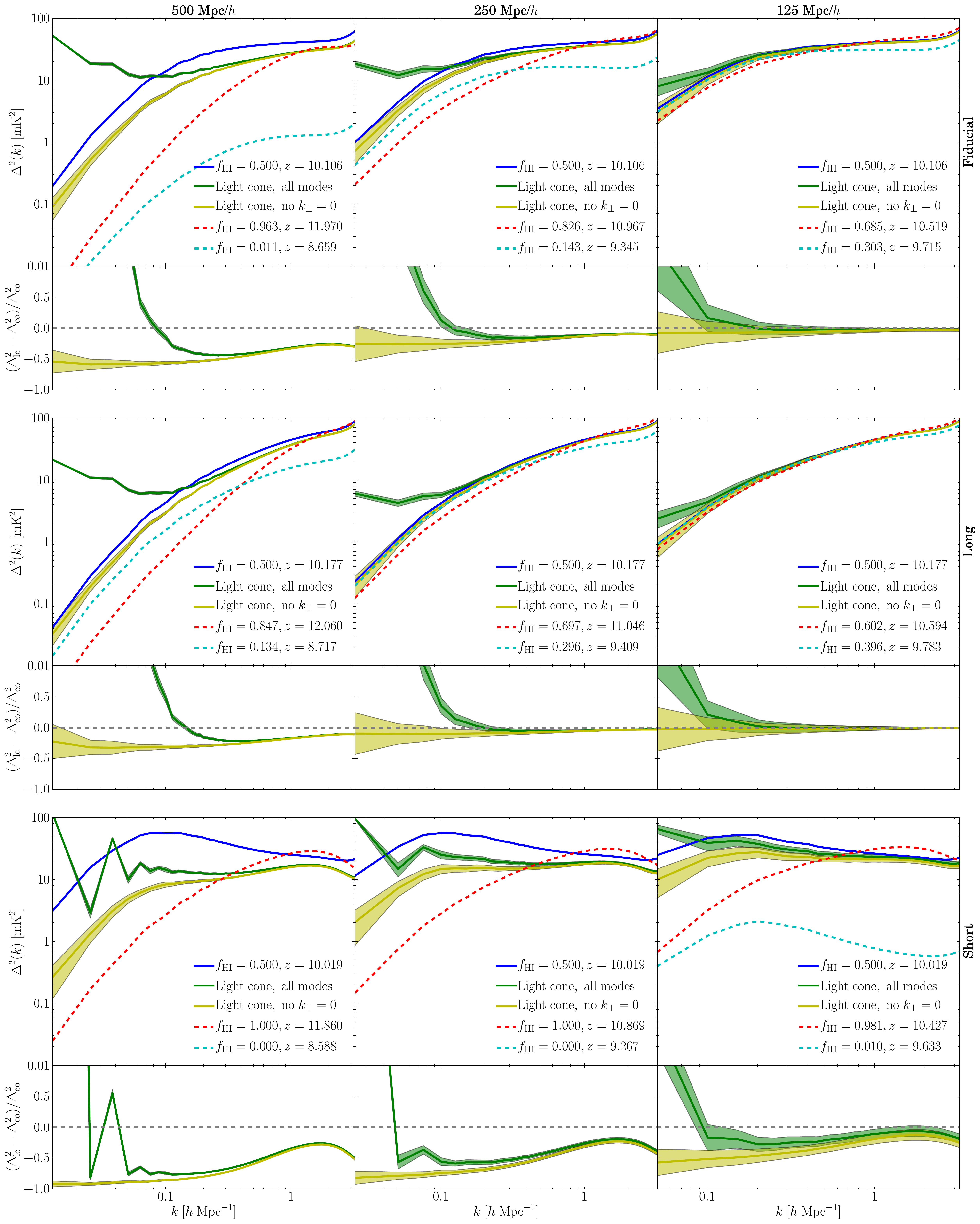}
  \caption{The same plot as in Figure \ref{fig:lc-ps-50}, but with all Fourier modes included. Note that the inclusion of the $k_\perp = 0$ mode still dramatically increases the power at small $k$-modes. On small scales, the inclusion of these modes do not change the signal significantly.}
  \label{fig:lc-ps-50-all}
\end{figure*}

Throughout the analysis, we computed different statistics both including and excluding modes where $k_\perp = 0$. In general, we find that removal of this mode causes the light cone case to appear similar to the coeval case. However, performing the analysis with $k_\perp = 0$ included has theoretical interest, since it explicitly demonstrates that the light cone effect shifts power from small scales to large scales. Plots similar to Figures~\ref{fig:lc-ps-50} and \ref{fig:lc-ps-fid}, but with all of the Fourier modes included, are shown in Figures~\ref{fig:lc-ps-75}, \ref{fig:lc-ps-50-all}, and \ref{fig:lc-ps-25}.

The inclusion of all Fourier modes in the analysis produces a signal that deviates by up to two orders of magnitude for large scales ($k \lesssim 0.05\ h$/Mpc). This deviation is with respect to both the light cone effect without these modes, and the coeval case. The dramatic increase in power at these scales is largely due to the combined change in neutral fraction during reionization. In other words, since there is a significant change in the mean temperature when examining large scales, there is much excess power on these scales. Note that the ringing in the case of the short reionization scenario is due to the sharp discontinuity between the front and back of the box.

This effect also introduces a strong anisotropy in the signal. When analyzing the signal using the power wedges analysis presented in \S\ref{sec:wedges}, we found that the modes parallel to the line of sight contributed about an order of magnitude more power than modes perpendicular, with all of this excess being due to the $k_\perp = 0$ mode.

\section{Additional Figures}
\label{sec:more-figs}
In addition to the plot presented in Figure~\ref{fig:lc-ps-50}, we also computed the light cone effect for all box sizes and reionization histories with $f_\mathrm{HI}=0.75, \, 0.25$. These plots are shown in Figures~\ref{fig:lc-ps-75} and \ref{fig:lc-ps-25}. Also, as mentioned in Appendix \ref{sec:exmodes}, these plots include all Fourier modes. As in the case of $f_\mathrm{HI} = 0.5$ in Figure~\ref{fig:lc-ps-50}, the light cone effect is still pronounced, though not quite as prominently. As before, the light cone effect is larger for bigger scales, and is most evident in the 500 Mpc/$h$ sub-box size. We conclude that regardless of the precise details of reionization, the light cone effect is an essential consideration for the 3D power spectrum of large volumes.

\begin{figure*}[p]
  \begin{minipage}[t]{0.325\hsize}
    \centering{\small $L = 250$ Mpc/$h$, Fiducial:}
  \end{minipage}
  \begin{minipage}[t]{0.305\hsize}
    \centering{\small $L = 250$ Mpc/$h$, Long:}
  \end{minipage}
  \begin{minipage}[t]{0.325\hsize}
    \centering{\small $L = 250$ Mpc/$h$, Short:}
  \end{minipage}
  \centering{
    \resizebox{0.325\hsize}{!}{\includegraphics{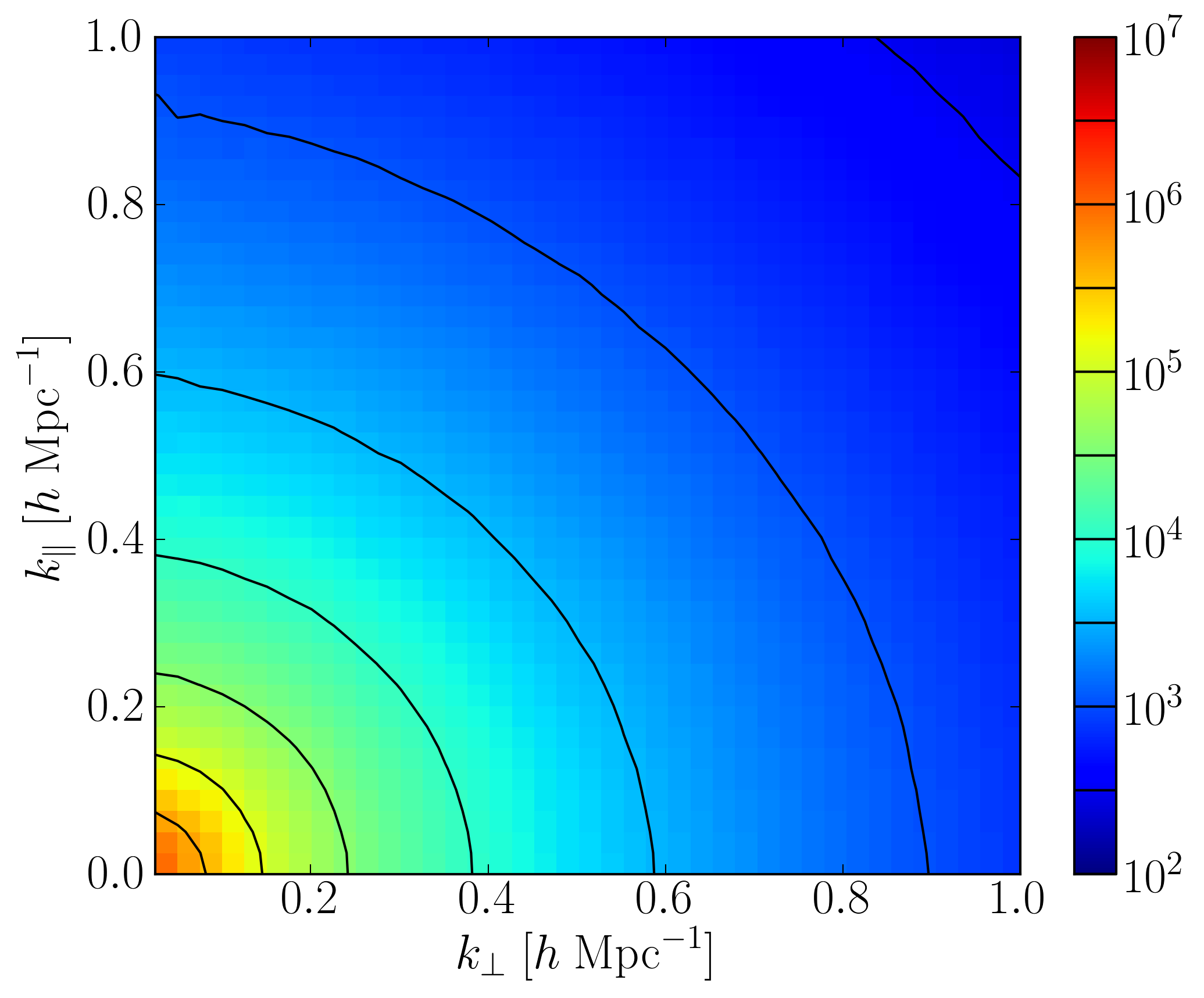}}%
    \hspace{2pt}%
    \resizebox{0.305\hsize}{!}{\includegraphics{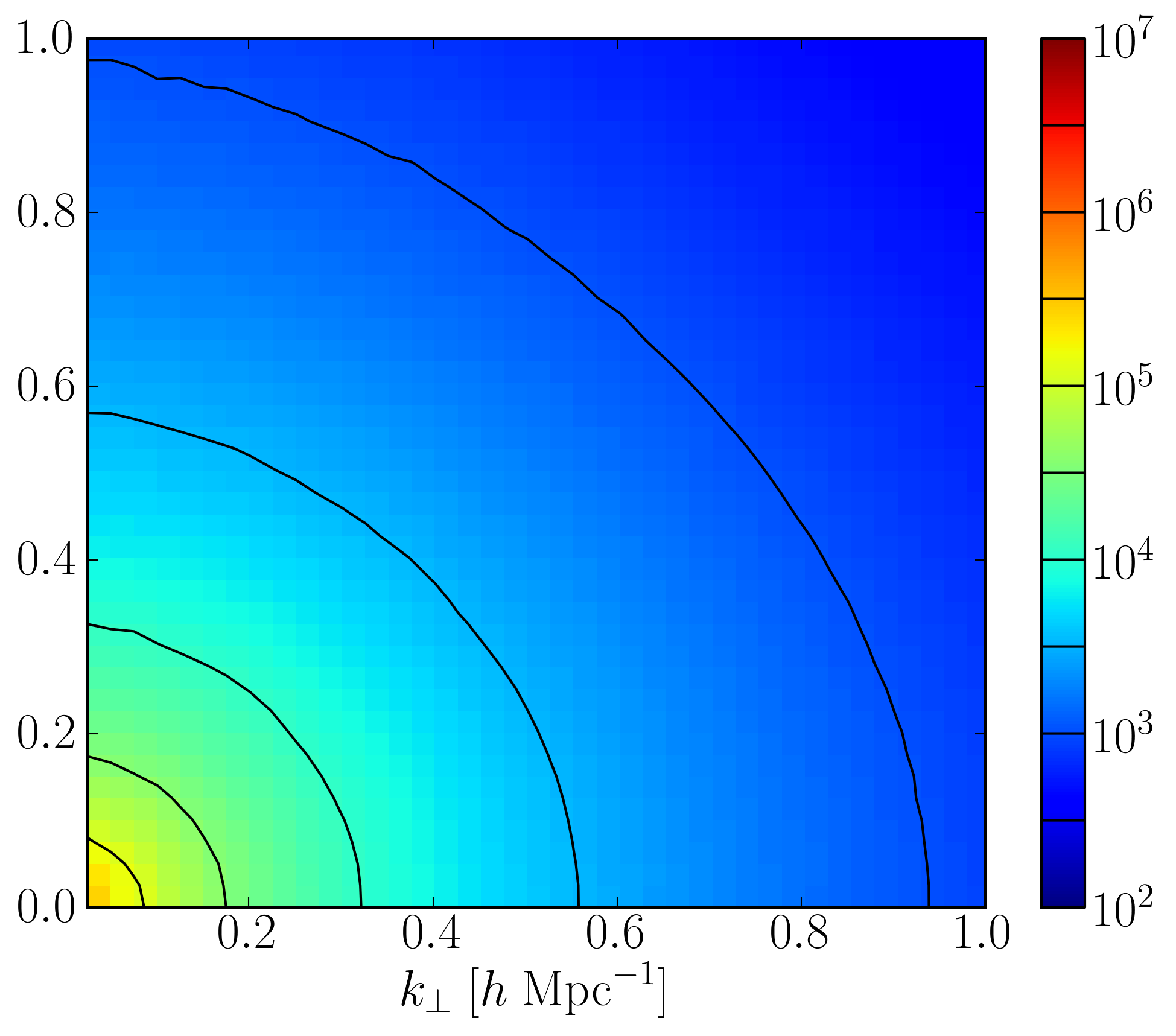}}%
    \hspace{2pt}%
    \resizebox{0.325\hsize}{!}{\includegraphics{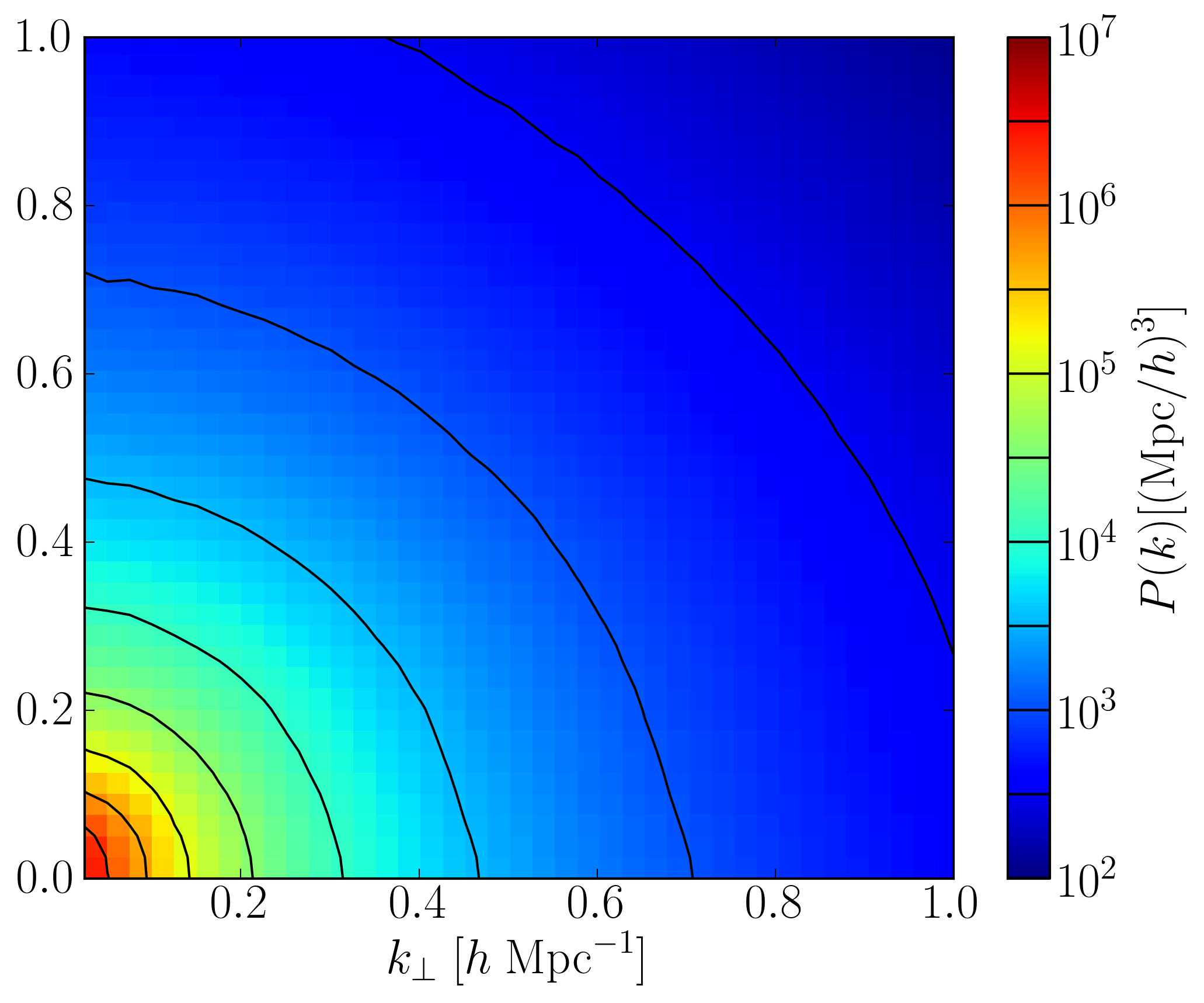}} }\\
  \begin{minipage}[t]{0.325\hsize}
    \centering{\small $L = 125$ Mpc/$h$, Fiducial:}
  \end{minipage}
  \begin{minipage}[t]{0.305\hsize}
    \centering{\small $L = 125$ Mpc/$h$, Long:}
  \end{minipage}
  \begin{minipage}[t]{0.325\hsize}
    \centering{\small $L = 125$ Mpc/$h$, Short:}
  \end{minipage}
  \centering{
    \resizebox{0.325\hsize}{!}{\includegraphics{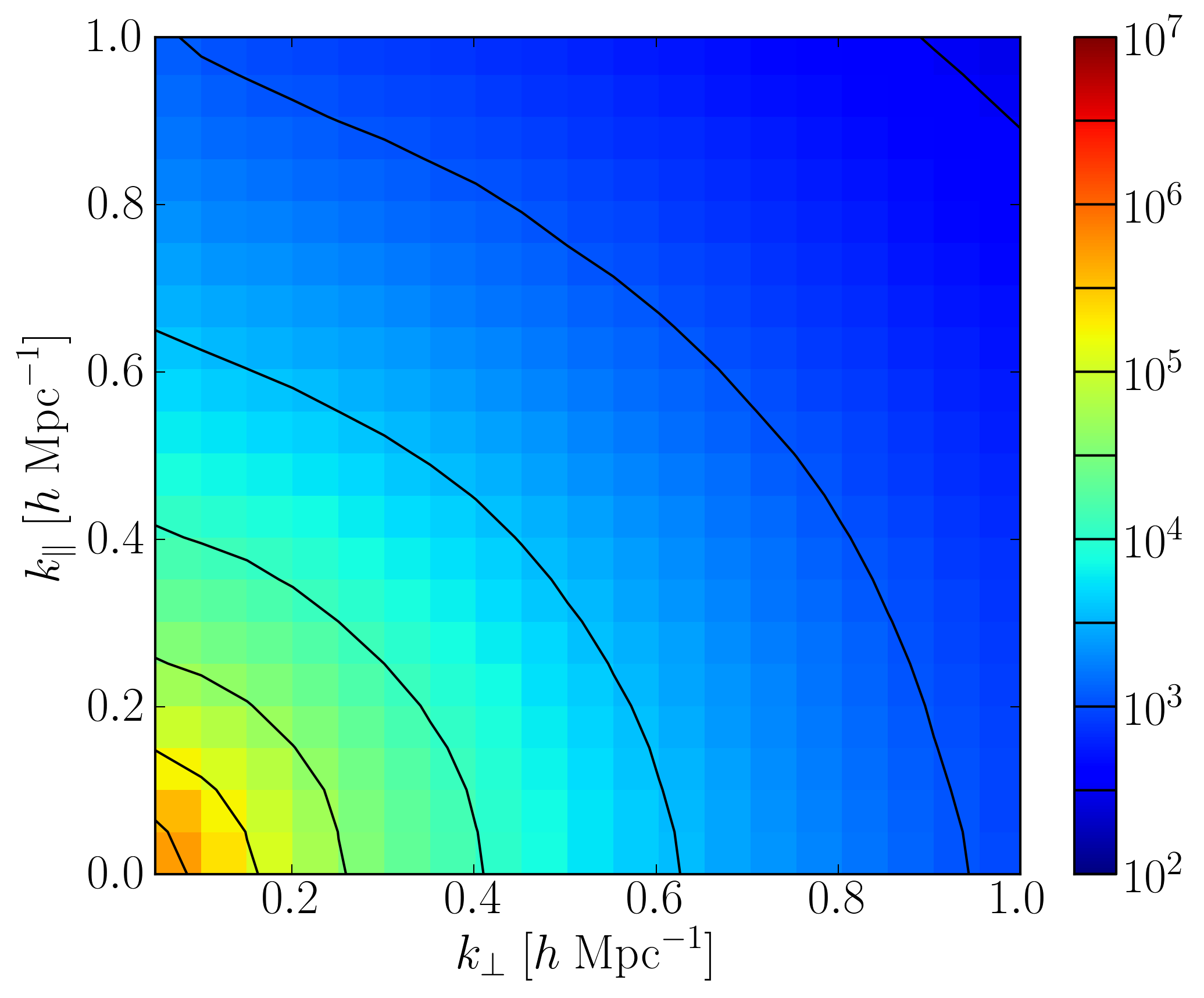}}%
    \hspace{2pt}%
    \resizebox{0.305\hsize}{!}{\includegraphics{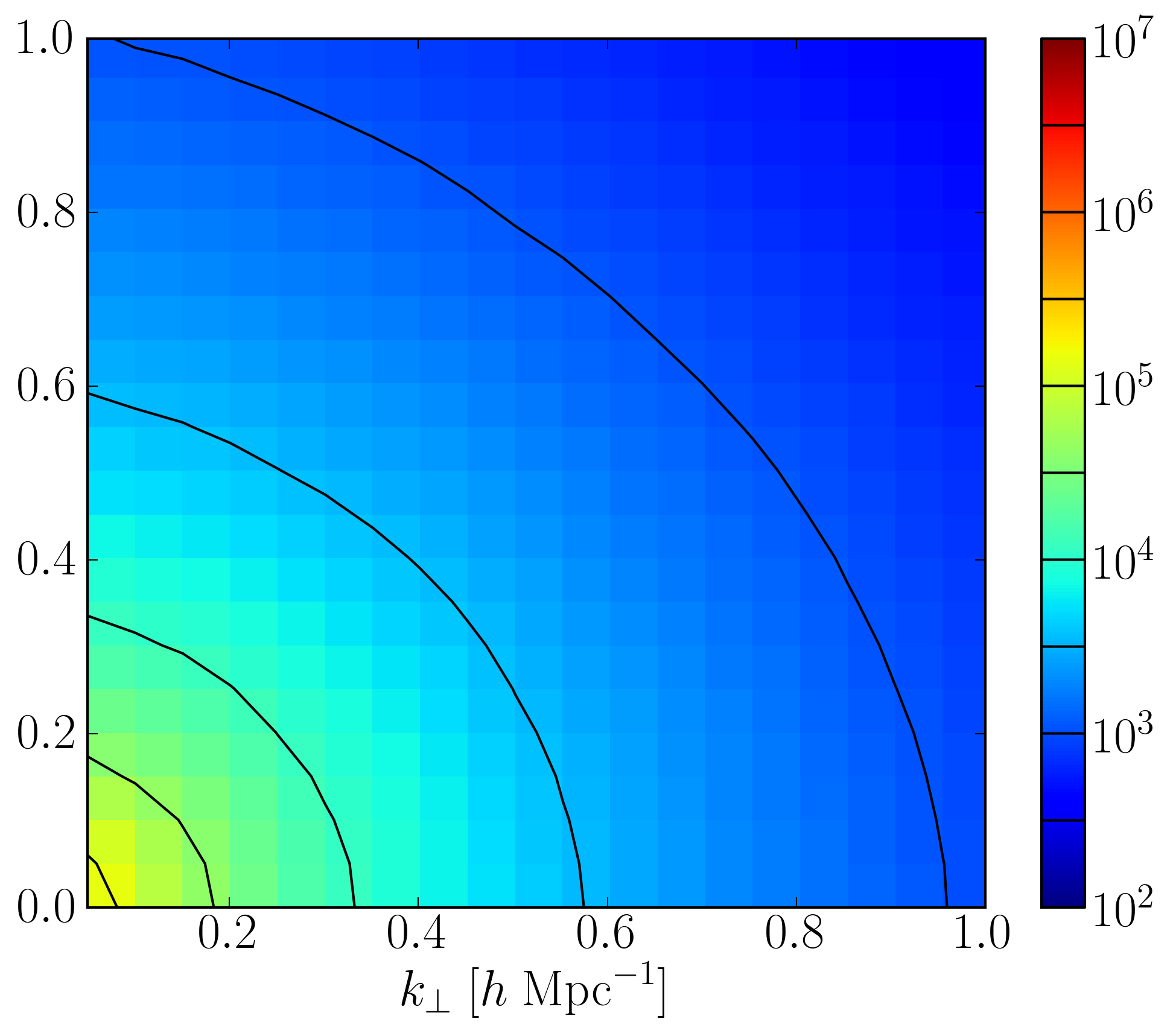}}%
    \hspace{2pt}%
    \resizebox{0.325\hsize}{!}{\includegraphics{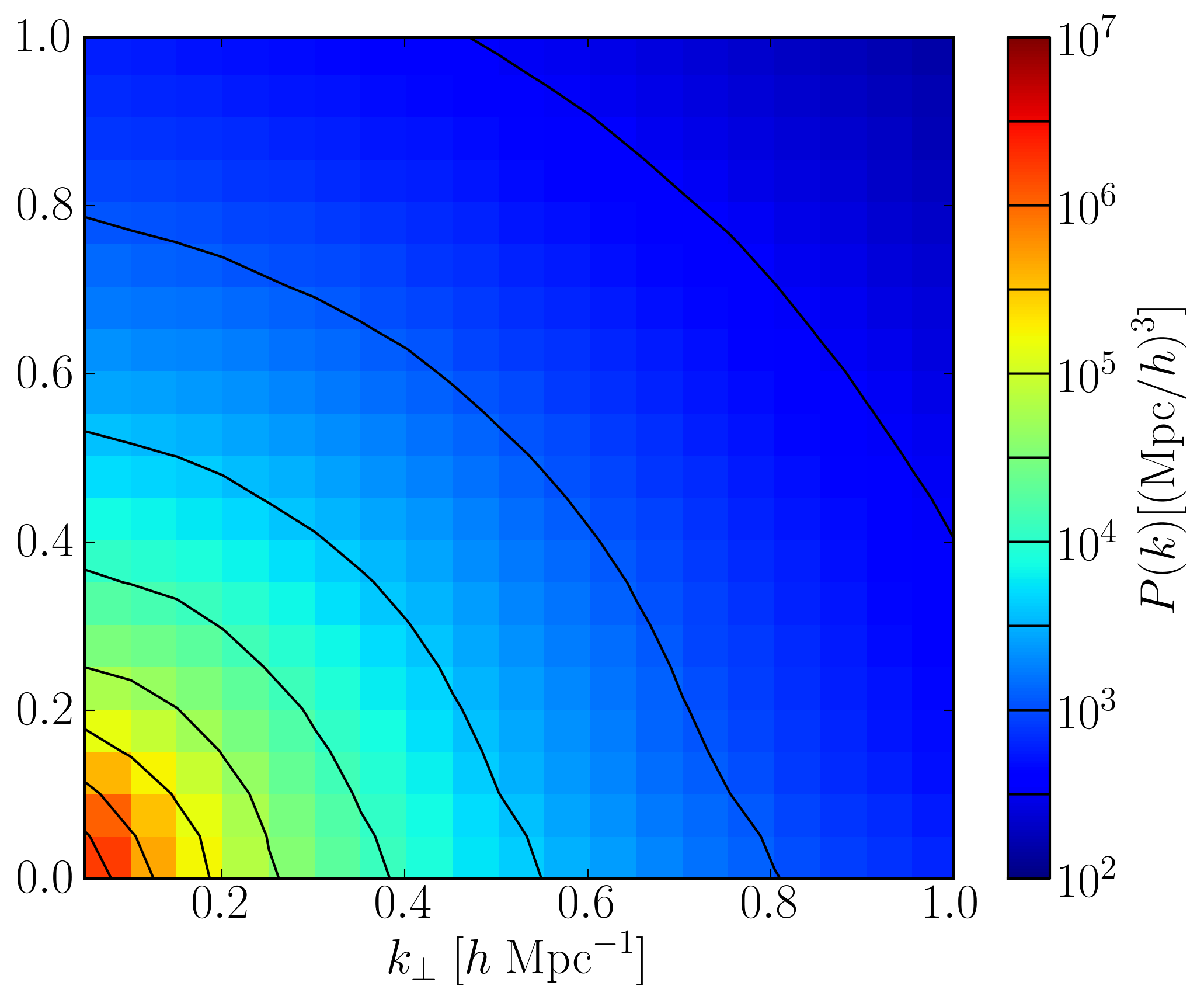}} }\\
  \caption{A comparison of the anisotropic power spectrum across different reionization histories and sub-box sizes. One interesting aspect in these plots is how the isopower lines are shaped: when the light cone effect is included, there is a change in the semi-circular contours. The central portion, near values where $k_\perp \sim k_\parallel$, has more power than regions where one component is much larger than the other. This is an interesting and subtle change in the contribution to the power introduced in the light cone.}
  \label{fig:2d-ps-all}
\end{figure*}

Figure~\ref{fig:2d-ps-all} shows the anisotropic power spectrum for the medium and small box sizes. At small scales, there is more power in $k_\perp > k_\parallel$ modes, with the exception of $k_\perp = 0$. Another interesting feature of these plots is how the shape of the isopower contours changes when the light cone effect is included. As discussed in \S\ref{sec:2dps}, the difference in the extent in redshift space and extent along the line of sky changes the amount of power for a given overall $k$. Also, as can be seen in Figure~\ref{fig:wedges-fid}, the anisotropy not including the $k_\perp = 0$ mode is greater for shorter reionization scenarios. For the smallest sub-box size, there is almost no anisotropy in most of the plot, because the extent in redshift space is small compared to the duration of reionization.


\begin{figure*}[p]
  \centering
  \includegraphics[width=\textwidth]{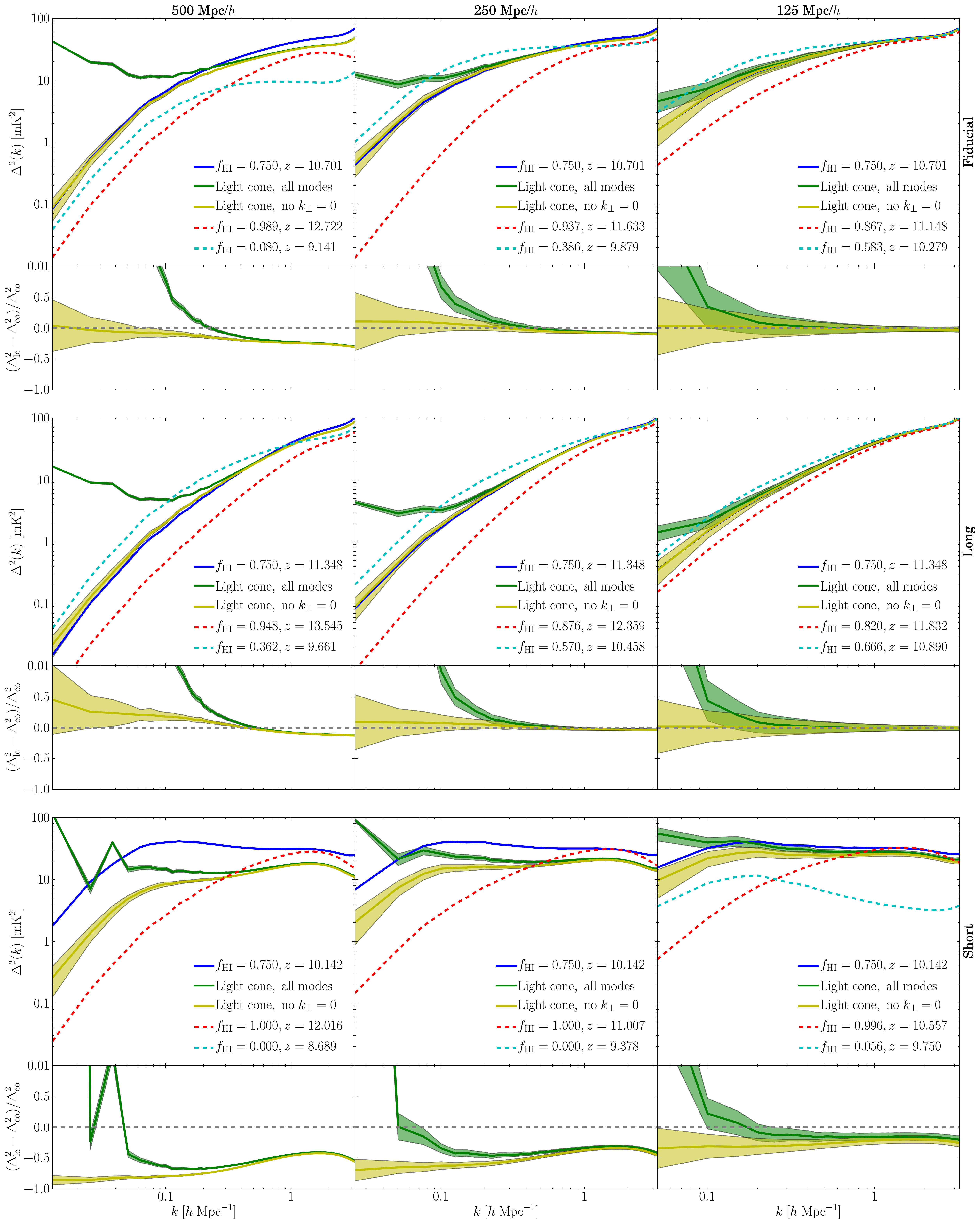}
  \caption{The same plot as in Figure~\ref{fig:lc-ps-50}, but at a 75\% ionization fraction and with all Fourier modes included. The inclusion of all Fourier modes produces a dramatic increase in the power spectrum, especially at small $k$-modes. (See the text in Appendix \ref{sec:exmodes} for more discussion.) We also find for the long and fiducial reionization scenarios that there is more power on large scales for the light cone than the coeval case. In general, the light cone effect at this neutral fraction is less pronounced, though still very significant. As in the main case of 50\% ionization, the effect is most noticeable for large box sizes. By extension, in the small sub-box case, the effect is still not very significant, as is the same for 50\% ionization fraction. One can also see that the shape of the power spectrum has changed dramatically in the case of short reionization.}
  \label{fig:lc-ps-75}
\end{figure*}

\begin{figure*}[p]
  \centering
  \includegraphics[width=\textwidth]{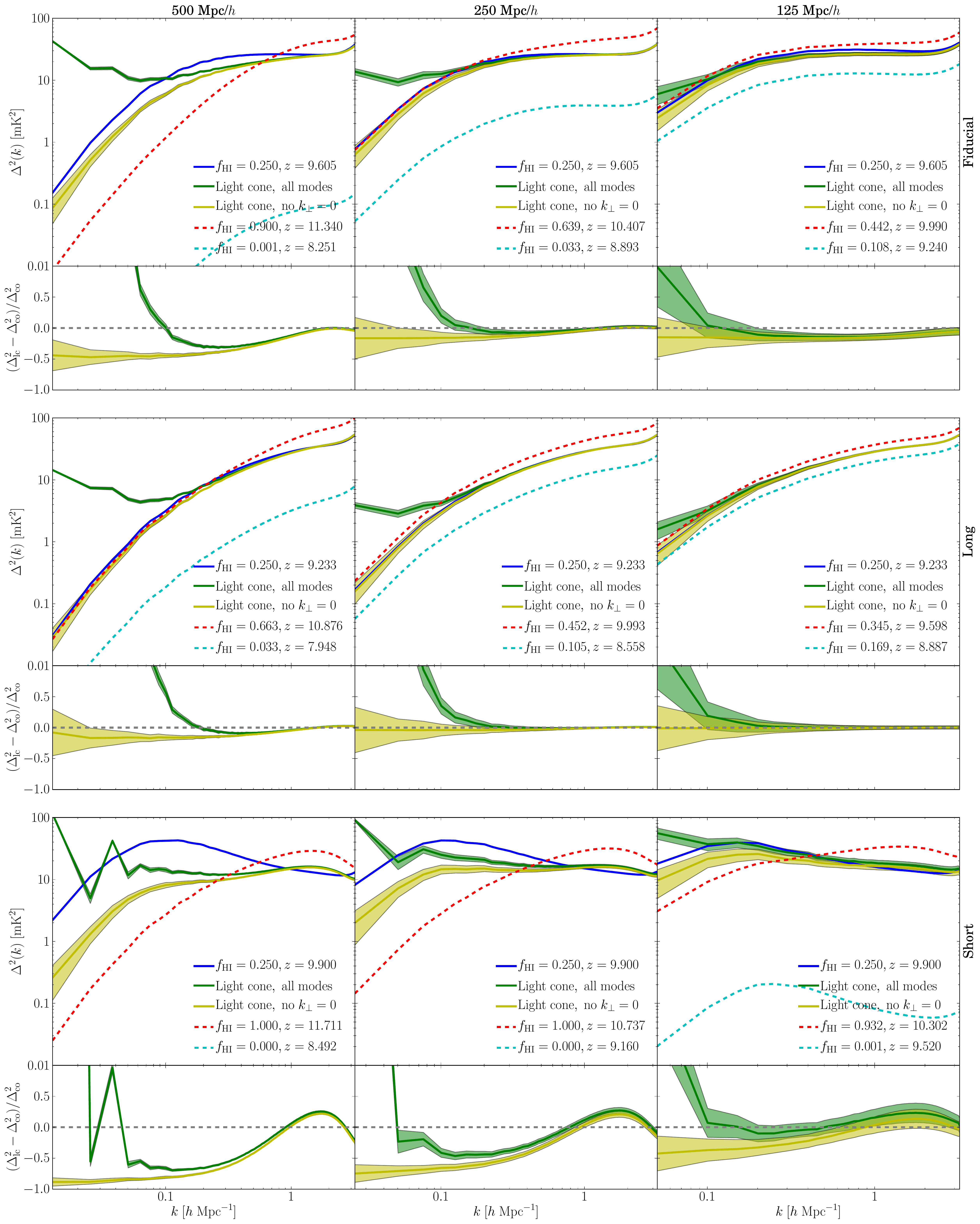}
  \caption{The same plot as in Figures~\ref{fig:lc-ps-50} and \ref{fig:lc-ps-75}, but at a 25\% ionization fraction. As with the case of a 75\% ionization fraction presented in Figure~\ref{fig:lc-ps-75}, in general the difference between the light cone and coeval cases is not as great as 50\% ionization. Nevertheless, is it still an important feature, and especially on the largest scales. One of the major implications is that the light cone effect is very important at large scales across a large ionization fraction range. Additionally, as in the coeval case, the light cone signal peaks at roughly a 50\% ionization fraction.}
  \label{fig:lc-ps-25}
\end{figure*}

\end{document}